\documentclass[preprint,pre,aps,amsfonts,amsmath,showpacs,superscriptaddress,nofootinbib, draft]{revtex4}
\usepackage{amsmath,amssymb,amscd}
\usepackage{bm,mathrsfs}
\usepackage{subeqnarray}
\begin{document}
\title{Quasi-Deterministic Properties of Random Gaussian Fields Constrained by a Large Quadratic Form}
\author{Philippe Mounaix}
\email{mounaix@cpht.polytechnique.fr}
\affiliation{Centre de Physique Th\'eorique, UMR 7644 du CNRS, Ecole
Polytechnique, 91128 Palaiseau Cedex, France.}
\date{\today}
\begin{abstract}
Completing the study initiated by Mounaix and Collet [J. Stat. Phys. {\bf 143}, 139-147 (2011)],  we investigate the realizations of a Gaussian random field in the limit where a given (general) quadratic form of the field is large. Concentration in $L^2$ and in probability is proved under mild conditions and the resulting quasi-deterministic behavior of the field is given. Applications to a large {\it local} quadratic form are considered in two specific cases. In particular, the quasi-deterministic structure of a Gaussian random flow with a large local helicity at some given point is determined explicitly.
\end{abstract}
\maketitle
%
\newtheorem{lemma}{Lemma}
\newtheorem{proposition}{Proposition}
\section{Introduction}\label{sec1}
This paper is devoted to the quasi-deterministic properties of a $n$-component Gaussian field on a finite domain of ${\mathbb R}^d$ in the limit where some quadratic form of the field is large. This is a generalization of the work by Mounaix and Collet (MC) which considered the case of a scalar field ($n=1$) constrained by a large $L^2$-norm\ \cite{MC} (see also\ \cite{MD}). While the primary motivation of MC was to get a better understanding of linear amplification in systems driven by the square of a Gaussian noise, this study may be of interest in a much wider range of physical situations, from smoothed laser beams to random surfaces and random flows. For instance, in the context of turbulent dynamo\ \cite{Mof}, the question arises whether the turbulent flow in the vicinity of a large helicity region has a quasi-deterministic structure or not. Since helicity is a quadratic form of the flow, our results make it possible to deal with this problem in any Kraichnan like model of advection, i.e. where the turbulent flow is approximated by a Gaussian random flow. We will briefly discuss this kind of applications in Sec.\ \ref{sec4b}.

In this paper, what we call `quasi-deterministic' is a generalization of the sense usually used in wave mechanics\ \cite{Boc}. Roughly, by `quasi-deterministic properties' we mean concentration properties by which the constrained field lives on a much smaller function space than the unconstrained one. More specifically, let $\varphi$ be a random field on a finite domain of ${\mathbb R}^d$ the realizations of which live in a separable Hilbert space $\mathscr{H}$. Let $\mathbb{P}_u$ denote the conditional probability knowing that some given real functional of $\varphi$ is greater than $u>0$. Let $\overline{\varphi}$ be a projection of $\varphi$ onto a finite, low-dimensional, subspace of $\mathscr{H}$. We say that $\varphi$ has a quasi-deterministic behavior (given by) $\overline{\varphi}$ if for every $\varepsilon >0$,
\begin{equation*}
\mathbb{P}_u\left(\left\|\frac{\varphi}{\|\varphi\|_2}
-\frac{\overline{\varphi}}{\|\overline{\varphi}\|_2}\right\|_2>\varepsilon\right)
\rightarrow 0,
\end{equation*}
as $u\rightarrow +\infty$. Quasi-deterministic behavior in the usual wave-mechanical sense is retrieved when the profile of $\overline{\varphi}$ is non random, which is the case if the function space on which $\overline{\varphi}$ lives is one-dimensional.

MC proved in\ \cite{MC} that if the considered functional of $\varphi$ is the square of its $L^2$-norm, $\|\varphi\|_2^2$, then $\varphi$ has a quasi-deterministic behavior with $\overline{\varphi}$ given by the orthogonal projection of $\varphi$ onto the fundamental eigenspace of its correlation operator (the dimension of which can be greater than one). By `fundamental eigenspace' we mean the eigenspace associated with the largest eigenvalue. Here we consider the general case of a quadratic form $\langle\varphi\vert\hat{O}\vert\varphi\rangle$, where $\hat{O}$ is a Hermitian operator on a distribution space containing $\mathscr{H}$. The problem considered by MC corresponds to the particular case $\hat{O}=\hat{1}$. Our main result states that, under mild conditions, (i) there is a quasi-deterministic behavior of $\varphi$ as $\vert\langle\varphi\vert\hat{O}\vert\varphi\rangle\vert\rightarrow +\infty$; and (ii) the corresponding $\overline{\varphi}$ is given by the contravariant component of $\varphi$ along the fundamental eigenspace of the restriction of $\pm\hat{C}\hat{O}$ to $\mathscr{D}(\hat{C}^{-1/2})$, where $\mathscr{D}(\hat{C}^{-1/2})\subset\mathscr{H}$ denotes the domain of $\hat{C}^{-1/2}$ and $\pm$ is the sign of $\langle\varphi\vert\hat{O}\vert\varphi\rangle$.

The outline of the paper is as follows. In Section\ \ref{sec2} we give some necessary definitions and we specify the class of $\varphi$ and $\hat{O}$ we consider. Section\ \ref{sec3} deals with the quasi deterministic behavior of $\varphi$ when $\vert\langle\varphi\vert\hat{O}\vert\varphi\rangle\vert$ is large. Finally, some possible applications are considered in Section\ \ref{sec4}. In particular, we determine the structure of a Gaussian random flow when its local helicity at some given point is large, which gives an answer to the question asked at the beginning of this Section (at least in its simplest version).
%
%
\section{Definitions}\label{sec2}
Let $\Lambda$ be a bounded subset of $\mathbb{R}^d$ and $\mathscr{H}=L^2(\Lambda)\otimes\mathbb{C}^N$ or $L^2(\Lambda)\otimes\mathbb{R}^N$, that is, the set of $N$-tuples $\lbrace\phi_1({\bm x}),\phi_2({\bm x}),\cdots ,\phi_N({\bm x})\rbrace$ of complex or real square-integrable functions on $\Lambda$. The inner product of $\mathscr{H}$ is defined by (using Dirac's bracket notation)
\begin{equation}\label{eq2.1}
\langle\psi\vert\phi\rangle = \sum_{m=1}^N\int_\Lambda\psi_m({\bm x})^\ast \phi_m({\bm x})\, d^dx.
\end{equation}
Let $\lbrace\vert\mu_n\rangle\rbrace$ ($n=1,2,\cdots$) be an orthonormal basis\footnote{such a basis exists as $\mathscr{H}$ is a separable Hilbert space} of $\mathscr{H}$, and $\hat{C}$ a positive, trace class, operator on $\mathscr{H}$ admitting the decomposition
\begin{equation}\label{eq2.2}
\hat{C}=\sum_n\mu_n\, \vert\mu_n\rangle\langle\mu_n\vert ,
\end{equation}
with $\mu_1\ge\mu_2\ge\cdots> 0$. $\hat{C}$ defines the covariance operator of a Gaussian measure on $\mathscr{H}$ with support $\mathscr{D}(\hat{C}^{-1/2})\subset\mathscr{H}$, the domain of $\hat{C}^{-1/2}$. The corresponding zero-mean Gaussian random field $\vert\varphi\rangle$ can be written as a generalized Karhunen-Loeve expansion\ \cite{Adl1}
\begin{equation}\label{eq2.3}
\vert\varphi\rangle =\sum_n t_n\hat{C}^{1/2}\vert\nu_n\rangle ,
\end{equation}
where $\lbrace\vert\nu_n\rangle\rbrace$ ($n=1,2,\cdots$) is an orthonormal basis of $\mathscr{H}$, and the $t_n$ are either i.i.d. complex Gaussian random variables with $\langle t_n\rangle =\langle t_n^2\rangle =0$ and  $\langle\vert t_n\vert^2\rangle =1$ (if $\mathscr{H}=L^2(\Lambda)\otimes\mathbb{C}^N$), or i.i.d. real Gaussian random variables with $\langle t_n\rangle =0$ and  $\langle t_n^2\rangle =1$ (if $\mathscr{H}=L^2(\Lambda)\otimes\mathbb{R}^N$).

Let $\hat{O}$ be a Hermitian operator on an appropriate distribution space containing $\mathscr{H}$. The choice of this distribution space (i.e. the choice of the appropriate rigged Hilbert space structure\ \cite{Gel}) depends on the considered problem. The class of $\hat{O}$ we consider is such that $\hat{M}\stackrel{d}{=}\hat{C}^{1/2}\hat{O}\hat{C}^{1/2}$ is a compact, Hermitian operator on $\mathscr{H}$. Write $\vert\lambda_n\rangle$ the eigenvector of $\hat{M}$ with eigenvalue $\lambda_n$. Since $\hat{M}$ is Hermitian, $\lbrace\vert\lambda_n\rangle\rbrace$ is an orthonormal basis of $\mathscr{H}$. Number the positive eigenvalues of $\hat{M}$ such that $\lambda_1\ge\lambda_2\ge\cdots$ and the negative ones such that $\lambda_{-1}\le\lambda_{-2}\le\cdots$. Let $g_{\pm 1}$ be the degeneracy of $\lambda_{\pm 1}$. Using\ (\ref{eq2.3}) with $\lbrace\vert\nu_n\rangle\rbrace =\lbrace\vert\lambda_n\rangle\rbrace$ we define
\begin{eqnarray}
&&\vert\overline{\varphi}_\pm\rangle =\sum_{n=1}^{g_{\pm 1}} t_{\pm n}\hat{C}^{1/2}\vert\lambda_{\pm n}\rangle ,\label{eq2.4}\\
&&\vert\delta\varphi_\pm\rangle =\vert\varphi\rangle -\vert\overline{\varphi}_\pm\rangle .\label{eq2.5}
\end{eqnarray}
Finally, for a fixed given value of $\pm$ such that $\vert\lambda_{\pm 1}\vert >0$ and every $u>0$, we write $d\mathbb{P}_u$ the conditional probability measure knowing that $\pm\langle\varphi\vert\hat{O}\vert\varphi\rangle >u$ and $\mathbb{E}_u$ the corresponding conditional expectation. Note that such a value of $\pm$ does exist since it follows from the inequalities $\sum_{n\ge 1}\vert t_{-n}\vert^2\lambda_{-n}\le\langle\varphi\vert\hat{O}\vert\varphi\rangle\le\sum_{n\ge 1}\vert t_{n}\vert^2\lambda_{n}$ and the requirement $\vert\langle\varphi\vert\hat{O}\vert\varphi\rangle\vert$ large that $\lambda_{-1}$ and $\lambda_1$ cannot be both zero.
%
%
\section{Quasi-deterministic behavior of $\bm{\varphi}$ for a large $\bm{\pm\langle\varphi\vert\hat{O}\vert\varphi\rangle}$}\label{sec3}
{\it A preliminary remark} : in the case of a real field, i.e. for $\mathscr{H}=L^2(\Lambda)\otimes\mathbb{R}^N$, one has $\langle\varphi\vert\hat{O}\vert\varphi\rangle = \langle\varphi\vert\hat{O}^S\vert\varphi\rangle$ and $\hat{O}$ needs not to be symmetric {\it a priori}, since it always appears through its symmetric part $\hat{O}^S$ only, which is the symmetric operator to be actually considered. So, in this section $\hat{O}$ always means $\hat{O}^S$ if $\mathscr{H}=L^2(\Lambda)\otimes\mathbb{R}^N$ and we drop the superscript $S$ for notation simplicity. (We will get back to the explicit notation $\hat{O}^S$ in Appendix\ \ref{app1}).

The quasi-deterministic behavior of $\varphi$ for a large $\pm\langle\varphi\vert\hat{O}\vert\varphi\rangle$ (with a fixed value of $\pm$) is embodied in the following proposition,
\begin{proposition}\label{prop1}
If $\hat{M}$ and $\hat{C}$ are trace class, then for every $\varepsilon >0$,
\begin{equation}\label{eq3.0}
\lim_{u\rightarrow +\infty}
\mathbb{P}_u\left(\left\|\frac{\varphi}{\|\varphi\|_2}
-\frac{\overline{\varphi}_\pm}{\|\overline{\varphi}_\pm\|_2}\right\|_2>\varepsilon\right)=0.
\end{equation}
\end{proposition}
For practical purposes it proves rather awkward to deal with operators like $\hat{M}$ (because of the presence of $\hat{C}^{1/2}$). This drawback can be remedied by the following proposition,
\begin{proposition}\label{prop2}
The spectrum of $\hat{M}$ is equal to the spectrum of the restriction of $\hat{C}\hat{O}$ to $\mathscr{D}(\hat{C}^{-1/2})$ with eigenstates $\lbrace\hat{C}^{1/2}\vert\lambda_n\rangle\rbrace$.
\end{proposition}
By this proposition it is possible to obtain the eigenvalues $\lambda_n$, as well as the vectors $\hat{C}^{1/2}\vert\lambda_{\pm n}\rangle$ appearing in the definition\ (\ref{eq2.4}) of $\vert\overline{\varphi}_\pm\rangle$, directly from the eigenvalue problem for $\hat{C}\hat{O}$, without dealing with $\hat{M}$ explicitely.

\bigskip
{\it Proof of Proposition\ \ref{prop1}.} We give the proof for ``$\pm =+$" and a complex random field, i.e. for $\mathscr{H}=L^2(\Lambda)\otimes\mathbb{C}^N$. The proof for ``$\pm =-$" is similar and the case of a real field, which follows the same line with technical differences, is given in Appendix\ \ref{app1}. Since the value of $\pm$ is fixed, we drop the subscript $\pm$ and write $\overline{\varphi}$ and $\delta\varphi$ for $\overline{\varphi}_\pm$ and $\delta\varphi_\pm$, respectively.

From the inequality $\|\varphi/\|\varphi\|_2-\overline{\varphi}/\|\overline{\varphi}\|_2\|_2\le 2\|\delta\varphi\|_2/\|\overline{\varphi}\|_2$ it follows that for every $\varepsilon >0$,
\begin{equation}\label{eq3.0b}
\mathbb{P}_u\left(\left\|\frac{\varphi}{\|\varphi\|_2}
-\frac{\overline{\varphi}}{\|\overline{\varphi}\|_2}\right\|_2>\varepsilon\right)\le
\mathbb{P}_u\left(\|\delta\varphi\|_2^2>\frac{\varepsilon^2}{4}
\|\overline{\varphi}\|_2^2\right).
\end{equation}
To prove\ (\ref{eq3.0}) we prove that under the conditions of proposition\ \ref{prop1},
\begin{equation}\label{eq3.1}
\lim_{u\rightarrow +\infty}\mathbb{P}_u\left(\|\delta\varphi\|_2^2>\varepsilon\|\overline{\varphi}\|_2^2\right)=0,
\end{equation}
and\ (\ref{eq3.0}) follows from\ (\ref{eq3.0b}) and\ (\ref{eq3.1}). For every $a>0$ one has
\begin{eqnarray}\label{eq3.2}
&&\mathbb{P}_u\left(\|\delta\varphi\|_2^2>\varepsilon\|\overline{\varphi}\|_2^2\right) \nonumber \\
&&=\mathbb{P}_u\left(\|\delta\varphi\|_2^2>\varepsilon\|\overline{\varphi}\|_2^2,\, \|\overline{\varphi}\|_2^2\ge a\right)
+\mathbb{P}_u\left(\|\delta\varphi\|_2^2>\varepsilon\|\overline{\varphi}\|_2^2,\, \|\overline{\varphi}\|_2^2<a\right)\nonumber \\
&&\le\mathbb{P}_u\left(\|\delta\varphi\|_2^2>\varepsilon a\right)+\mathbb{P}_u\left(\|\overline{\varphi}\|_2^2<a\right).
\end{eqnarray}
To estimate the two conditional probabilities on the right-hand side of\ (\ref{eq3.2}) we first need to bound $\mathbb{P}(\langle\varphi\vert\hat{O}\vert\varphi\rangle >u)$ from below. Let $\rho(v)$ denote the probability distribution function (pdf) of $V=\sum_{\lbrace 1\le n\le g_1\rbrace\cup\lbrace n<0\rbrace}\lambda_n\vert t_n\vert^2$. Writing $\rho(v)$ as the inverse Fourier transform of the characteristic function of $V$, one has the integral representation
\begin{equation}\label{eq3.3}
\rho(v)=\int_{-\infty}^{+\infty}\frac{\exp(-ikv)}{(1-ik\lambda_1)^{g_1}}\, \prod_{n>0}\frac{1}{(1-ik\lambda_{-n})}\, \frac{dk}{2\pi}.
\end{equation}
The existence of the product on the right-hand side of\ (\ref{eq3.3}) is ensured by $\hat{M}$ being trace class. For $v>0$, only the pole at $k=-i/\lambda_1$ contributes to\ (\ref{eq3.3}) and one gets
\begin{equation}\label{eq3.4}
\rho(v)=P_{g_1 -1}(v)\exp\left(-\frac{v}{\lambda_1}\right),
\end{equation}
where $P_{g_1 -1}(v)$ is a polynomial of degree $g_1 -1$ (Note that $\hat{M}$ being trace class ensures that the coefficients of $P_{g_1 -1}(v)$ exist). For large $v$ the polynomial $P_{g_1 -1}(v)$ is dominated by its highest order term and one has
\begin{equation}\label{eq3.5}
\rho(v)\sim\frac{1}{(g_1-1)!}\prod_{n>0}\frac{1}{(1-\lambda_{-n}/\lambda_1)}\, \left(\frac{v}{\lambda_1}\right)^{g_1-1}
\frac{\exp(-v/\lambda_1)}{\lambda_1}\ \ \ \ (v\rightarrow +\infty),
\end{equation}
from which it follows that for every $0<\alpha <1$, there is $v_0 >0$ such that for every $v>v_0$
\begin{equation}\label{eq3.6}
\rho(v)\ge\frac{(1-\alpha)}{(g_1-1)!}\prod_{n>0}\frac{1}{(1-\lambda_{-n}/\lambda_1)}\, \left(\frac{v}{\lambda_1}\right)^{g_1-1}
\frac{\exp(-v/\lambda_1)}{\lambda_1}.
\end{equation}
From\ (\ref{eq3.6}) and the lower bound
\begin{eqnarray}\label{eq3.7}
\mathbb{P}\left(\langle\varphi\vert\hat{O}\vert\varphi\rangle >u\right)&\equiv&
\mathbb{P}\left(\sum_n \lambda_n\vert t_n\vert^2 >u\right) \nonumber \\
&\ge&\mathbb{P}\left(\sum_{\lbrace 1\le n\le g_1\rbrace\cup\lbrace n<0\rbrace}\lambda_n\vert t_n\vert^2 >u\right) \nonumber \\
&=&\int_u^{+\infty}\rho(v)\, dv,
\end{eqnarray}
one finds that for $u$ large enough (i.e. $u>v_0$),
\begin{equation}\label{eq3.8}
\mathbb{P}\left(\langle\varphi\vert\hat{O}\vert\varphi\rangle >u\right)\ge C_1(\alpha)\, u^{g_1-1}
\exp\left(-\frac{u}{\lambda_1}\right),
\end{equation}
with
\begin{equation}\label{eq3.9}
C_1(\alpha) = \frac{(1-\alpha)}{(g_1-1)!\, \lambda_1^{g_1-1}}
\prod_{n>0}\frac{1}{(1-\lambda_{-n}/\lambda_1)}.
\end{equation}
First, we estimate the conditional probabilities $\mathbb{P}_u\left(\|\overline{\varphi}\|_2^2<a\right)$. One has
\begin{eqnarray}\label{eq3.10}
&&\mathbb{P}\left(\|\overline{\varphi}\|_2^2 <a,\, \langle\varphi\vert\hat{O}\vert\varphi\rangle >u\right) =
\mathbb{P}\left(\sum_{i,j=1}^{g_1}\langle\lambda_i\vert\hat{C}\vert\lambda_j\rangle t_i^\ast t_j <a,\, 
\sum_i\lambda_i \vert t_i\vert^2 >u\right) \nonumber \\
&&\le\mathbb{P}\left(\sum_{i,j=1}^{g_1}\langle\lambda_i\vert\hat{C}\vert\lambda_j\rangle t_i^\ast t_j <a,\, 
\sum_{i\ge 1}\lambda_i \vert t_i\vert^2 >u\right).
\end{eqnarray}
The matrix $\langle\lambda_i\vert\hat{C}\vert\lambda_j\rangle$ is a $g_1\times g_1$ positive definite Hermitian matrix. Let $\tilde{\mu}_1\ge\tilde{\mu}_2\ge\cdots\ge\tilde{\mu}_{g_1}$ denote its eigenvalues and $\lbrace\vert\tilde{\mu}_i\rangle\rbrace$ the corresponding orthonormal basis of eigenvectors. For every realization of the $t_i$ one has
\begin{equation}\label{eq3.11}
\sum_{i,j=1}^{g_1}\langle\lambda_i\vert\hat{C}\vert\lambda_j\rangle t_i^\ast t_j =\sum_{n=1}^{g_1}\tilde{\mu}_n \vert\tilde{t}_n\vert^2,
\end{equation}
where
\begin{equation}\label{eq3.12}
\tilde{t}_n=\sum_{i=1}^{g_1}\langle\tilde{\mu}_n\vert\lambda_i\rangle t_i.
\end{equation}
From\ (\ref{eq3.12}) one finds that the $\tilde{t}_i$ have the same statistical properties as the $t_i$ (i.e. they are i.i.d. complex Gaussian random variables with $\langle\tilde{t}_i\rangle =\langle\tilde{t}_i^2\rangle =0$ and  $\langle\vert\tilde{t}_i\vert^2\rangle =1$), with
\begin{equation}\label{eq3.13}
\sum_{i=1}^{g_1}\vert t_i\vert^2 =\sum_{i=1}^{g_1}\vert\tilde{t}_i\vert^2.
\end{equation}
Using\ (\ref{eq3.12}) and\ (\ref{eq3.13}) on the right-hand side of\ (\ref{eq3.10}) and dropping the tilde (because the $\tilde{t}_i$ and the $t_i$ have the same statistical properties), one obtains
\begin{equation}\label{eq3.14}
\mathbb{P}\left(\|\overline{\varphi}\|_2^2 <a,\, \langle\varphi\vert\hat{O}\vert\varphi\rangle >u\right)\le
\mathbb{P}\left(\sum_{i=1}^{g_1}\tilde{\mu}_i \vert t_i\vert^2 <a,\, \sum_{i\ge 1}\lambda_i \vert t_i\vert^2 >u\right).
\end{equation}
It follows from $\mu_n>0$ for every $n$ and\ (\ref{eq2.2}) that $\tilde{\mu}_i =\langle\tilde{\mu}_i\vert\hat{C}\vert\tilde{\mu}_i\rangle >0$ for every $i$, and from $\hat{M}$ being trace class that $g_1<+\infty$. Thus,
\begin{equation}\label{eq.3.15}
\tilde{\mu}_{min}=\inf_{1\le i\le g_1}\left\lbrace\tilde{\mu}_i\, :\, \tilde{\mu}_i>0\right\rbrace =\tilde{\mu}_{g_1}>0,
\end{equation}
and\ (\ref{eq3.14}) is bounded above by
\begin{eqnarray}\label{eq3.16}
&&\mathbb{P}\left(\|\overline{\varphi}\|_2^2 <a,\, \langle\varphi\vert\hat{O}\vert\varphi\rangle >u\right)\le
\mathbb{P}\left(\tilde{\mu}_{min}\sum_{i=1}^{g_1}\vert t_i\vert^2 <a,\, \sum_{i\ge 1}\lambda_i \vert t_i\vert^2 >u\right) \nonumber \\
&&=\int_{x=0}^{+\infty}\mathbb{P}\left(\frac{u-x}{\lambda_1} <\sum_{i=1}^{g_1}\vert t_i\vert^2 <\frac{a}{\tilde{\mu}_{min}}\right)\, 
d\mathbb{P}\left(\sum_{i>g_1}\lambda_i\vert t_i\vert^2 =x\right) \nonumber \\
&&=\int_{x=u-\lambda_1 a/\tilde{\mu}_{min}}^{+\infty}\mathbb{P}\left(\frac{u-x}{\lambda_1} <\sum_{i=1}^{g_1}\vert t_i\vert^2 <\frac{a}{\tilde{\mu}_{min}}\right)\, d\mathbb{P}\left(\sum_{i>g_1}\lambda_i\vert t_i\vert^2 =x\right) \nonumber \\
&&\le\int_{x=u-\lambda_1 a/\tilde{\mu}_{min}}^{+\infty}d\mathbb{P}\left(\sum_{i>g_1}\lambda_i\vert t_i\vert^2 =x\right) \nonumber \\
&&=\mathbb{P}\left(\sum_{i>g_1}\lambda_i\vert t_i\vert^2 >u-\frac{\lambda_1 a}{\tilde{\mu}_{min}}\right),
\end{eqnarray}
where we have used the statistical independence of the $t_i$ (second line), and the fact that the probability in the integrand vanishes identically for $x<u-\lambda_1 a/\tilde{\mu}_{min}$ (third line). Now, by exponential Markov inequality, one has for every positive $c<1/\lambda_{g_1+1}$,
\begin{eqnarray*}
\mathbb{P}\left(\sum_{i>g_1}\lambda_i\vert t_i\vert^2 >u-\frac{\lambda_1 a}{\tilde{\mu}_{min}}\right)&\le&
{\rm e}^{-c(u-\lambda_1 a/\tilde{\mu}_{min})}\mathbb{E}\left\lbrack\exp\left(c\sum_{i>g_1}\lambda_i\vert t_i\vert^2\right)\right\rbrack \nonumber \\
&=&{\rm e}^{-c(u-\lambda_1 a/\tilde{\mu}_{min})}\prod_{i>g_1}\frac{1}{(1-c\lambda_i)},
\end{eqnarray*}
and by taking $c=(\lambda_1^{-1}+\lambda_{g_1+1}^{-1})/2$, one gets
\begin{equation}\label{eq3.17}
\mathbb{P}\left(\|\overline{\varphi}\|_2^2 <a,\, \langle\varphi\vert\hat{O}\vert\varphi\rangle >u\right)\le
C_2(a) \exp\left\lbrack -\left(\frac{1}{\lambda_1}+\frac{1}{\lambda_{g_1+1}}\right)\frac{u}{2}\right\rbrack ,
\end{equation}
with
\begin{equation}\label{eq3.18}
C_2(a)=\exp\left\lbrack\left(1+\frac{\lambda_1}{\lambda_{g_1+1}}\right)\frac{a}{2\tilde{\mu}_{min}}\right\rbrack
\prod_{i>g_1}\frac{1}{\lbrack 1-(\lambda_1^{-1}+\lambda_{g_1+1}^{-1})\lambda_i/2\rbrack}.
\end{equation}
The existence of the product on the right-hand side of\ (\ref{eq3.18}) is ensured by $\hat{M}$ being trace class. Finally, it follows from the two estimates\ (\ref{eq3.8}) and\ (\ref{eq3.17}) that for $u$ large enough $\mathbb{P}_u\left(\|\overline{\varphi}\|_2^2<a\right)$ is bounded above by
\begin{eqnarray}\label{eq3.19}
\mathbb{P}_u\left(\|\overline{\varphi}\|_2^2<a\right) &\equiv&
\frac{\mathbb{P}\left(\|\overline{\varphi}\|_2^2 <a,\, \langle\varphi\vert\hat{O}\vert\varphi\rangle >u\right)}
{\mathbb{P}\left(\langle\varphi\vert\hat{O}\vert\varphi\rangle >u\right)} \nonumber \\
&\le&\frac{C_2(a)}{C_1(\alpha)}\, \frac{1}{u^{g_1-1}}\, 
\exp\left\lbrack -\left(\frac{1}{\lambda_{g_1+1}}-\frac{1}{\lambda_1}\right)\frac{u}{2}\right\rbrack .
\end{eqnarray}
We now estimate the conditional probabilities $\mathbb{P}_u\left(\|\delta\varphi\|_2^2>\varepsilon a\right)$. First, we prove that there exists $c>0$ such that $\mathbb{E}_u\left\lbrack\exp\left( c\|\delta\varphi\|_2^2\right)\right\rbrack <+\infty$. One has
\begin{eqnarray}\label{eq3.20}
&&d\mathbb{P}\left(\lbrace t_{i\notin\lbrack 1,g_1\rbrack}\rbrace ,\, \sum_i \lambda_i\vert t_i\vert^2 >u\right)
\le d\mathbb{P}\left(\lbrace t_{i\notin\lbrack 1,g_1\rbrack}\rbrace ,\, \sum_{i\ge 1} \lambda_i\vert t_i\vert^2 >u\right) \nonumber \\
&&=\mathbb{P}\left(\sum_{i=1}^{g_1}\vert t_i\vert^2 >\frac{u}{\lambda_1}-\sum_{i>g_1}\frac{\lambda_i}{\lambda_1}\vert t_i\vert^2\right)
\, \prod_{i\notin\lbrack 1,g_1\rbrack}\frac{{\rm e}^{-\vert t_i\vert^2}}{\pi}\, d^2t_i,
\end{eqnarray}
and
\begin{eqnarray}\label{eq3.21}
&&\mathbb{P}\left(\sum_{i=1}^{g_1}\vert t_i\vert^2 >\frac{u}{\lambda_1}-\sum_{i>g_1}\frac{\lambda_i}{\lambda_1}\vert t_i\vert^2\right) \nonumber \\
&&=\int_{(u-\sum_{i>g_1}\lambda_i\vert t_i\vert^2)/\lambda_1}^{+\infty}\frac{H(v)v^{g_1-1}}{(g_1-1)!}\, {\rm e}^{-v}\, dv \\
&&\le\frac{1}{(g_1-1)!}\int_{u/\lambda_1}^{+\infty}v^{g_1-1}\, {\rm e}^{-v}\, dv\, \prod_{i>g_1}{\rm e}^{\lambda_i\vert t_i\vert^2/\lambda_1}, \nonumber
\end{eqnarray}
where $H(v)$ is the Heaviside step function. The third line of\ (\ref{eq3.21}) is obtained by making the change of variable $v\rightarrow v-\sum_{i>g_1}\lambda_i\vert t_i\vert^2)/\lambda_1$ in the second line, and by using the fact that $H(v)v^{g_1-1}$ is an increasing function of $v$. For large $u$ one has
\begin{equation}\label{eq3.22}
\int_{u/\lambda_1}^{+\infty}v^{g_1-1}\, {\rm e}^{-v}\, dv
\sim\left(\frac{u}{\lambda_1}\right)^{g_1-1}\, {\rm e}^{-u/\lambda_1}\ \ \ \ (u\rightarrow +\infty),
\end{equation}
from which it follows that for every $\alpha >0$, there is $v_1 >0$ such that for every $u>v_1$
\begin{equation}\label{eq3.23}
\mathbb{P}\left(\sum_{i=1}^{g_1}\vert t_i\vert^2 >\frac{u}{\lambda_1}-\sum_{i>g_1}\frac{\lambda_i}{\lambda_1}\vert t_i\vert^2\right)\le
\frac{1+\alpha}{(g_1-1)!}\left(\frac{u}{\lambda_1}\right)^{g_1-1}\, {\rm e}^{-u/\lambda_1}
\, \prod_{i>g_1}{\rm e}^{\lambda_i\vert t_i\vert^2/\lambda_1}.
\end{equation}
Taking $0<\alpha <1$, it follows from\ (\ref{eq3.8}),\ (\ref{eq3.20}), and\ (\ref{eq3.23}) that for $u$ large enough (i.e. $u>\max\lbrace v_0,v_1\rbrace$)
\begin{equation}\label{eq3.24}
d\mathbb{P}_u\left(\lbrace t_{i\notin\lbrack 1,g_1\rbrack}\rbrace\right)\le C_3(\alpha)
\, \left(\prod_{i<0}\frac{{\rm e}^{-\vert t_i\vert^2}}{\pi}\, d^2t_i\right)
\, \left(\prod_{i>g_1}\frac{{\rm e}^{-(1-\lambda_i/\lambda_1)\vert t_i\vert^2}}{\pi}\, d^2t_i\right),
\end{equation}
with
\begin{equation}\label{eq3.25}
C_3(\alpha)=\frac{(1+\alpha)}{(1-\alpha)}\, \prod_{n>0}\left(1-\frac{\lambda_{-n}}{\lambda_1}\right).
\end{equation}
The existence of the product on the right-hand side of\ (\ref{eq3.25}) is ensured by $\hat{M}$ being trace class. Using\ (\ref{eq3.24}) to estimate $\mathbb{E}_u\left\lbrack\exp\left( c\|\delta\varphi\|_2^2\right)\right\rbrack$, one finds
\begin{equation}\label{eq3.26}
\mathbb{E}_u\left\lbrack\exp\left( c\|\delta\varphi\|_2^2\right)\right\rbrack\le C_3(\alpha)C_4(c),
\end{equation}
where $C_4(c)$ is an infinite product of Gaussian integrals the existence of which is ensured if the matrix
\begin{equation*}
{\rm diag}\left(\min\left\lbrace 1,\, 1-\lambda_i/\lambda_1\right\rbrace\right)
-c\, \langle\lambda_i\vert\hat{C}\vert\lambda_j\rangle ,
\end{equation*}
is (strictly) positive definite, and the matrix
\begin{equation*}
{\rm diag}\left(\max\left\lbrace 0,\, \lambda_i/\lambda_1\right\rbrace\right)
+c\, \langle\lambda_i\vert\hat{C}\vert\lambda_j\rangle ,
\end{equation*}
is trace class, with $i$ and $j$ $\notin\lbrack 1,\, g_1\rbrack$. The latter condition is fulfilled by $\hat{M}$ and $\hat{C}$ being trace class. The former one requires $0<c<(1-\lambda_{g_1+1}/\lambda_1)/\mu_1$, which follows from $\|\hat{C}\| = \mu_1 <+\infty$ and $\min\left\lbrace 1,\, 1-\lambda_i/\lambda_1\right\rbrace\ge 1-\lambda_{g_1+1}/\lambda_1>0$ for $i\notin\lbrack 1,\, g_1\rbrack$. Thus, there exists $c>0$ such that $C_4(c)<+\infty$, and by exponential Markov inequality and Eq.\ (\ref{eq3.26}) one finds that for $u$ large enough
\begin{eqnarray}\label{eq3.27}
\mathbb{P}_u\left(\|\delta\varphi\|_2^2>\varepsilon a\right)&\le& {\rm e}^{-\varepsilon ca}
\mathbb{E}_u\left\lbrack\exp\left( c\|\delta\varphi\|_2^2\right)\right\rbrack \nonumber \\
&\le&C_3(\alpha)C_4(c)\, {\rm e}^{-\varepsilon ca}.
\end{eqnarray}
Finally, using the estimates\ (\ref{eq3.19}) and\ (\ref{eq3.27}) on the right-hand side of\ (\ref{eq3.2}) and taking the limit $u\rightarrow +\infty$, one obtains
\begin{equation}\label{eq3.28}
\lim_{u\rightarrow +\infty}\mathbb{P}_u\left(\|\delta\varphi\|_2^2>\varepsilon\|\overline{\varphi}\|_2^2\right)\le
 C_3(\alpha)C_4(c)\, {\rm e}^{-\varepsilon ca},
\end{equation}
and $a>0$ being arbitrary completes the proof of Proposition\ \ref{prop1}. [The right-hand side of\ (\ref{eq3.28}) can be made arbitrarily small.] $\square$

\bigskip
{\it Proof of Proposition\ \ref{prop2}.} Let $\vert\lambda\rangle\in\mathscr{H}$ be a normalized eigenstate of $\hat{M}$ with eigenvalue $\lambda$. One has
\begin{equation}\label{eq3.29}
\hat{M}\vert\lambda\rangle =\hat{C}^{1/2}\hat{O}\hat{C}^{1/2}\vert\lambda\rangle =\lambda\vert\lambda\rangle .
\end{equation}
Applying $\hat{C}^{1/2}$ to both sides of\ (\ref{eq3.29}) gives
\begin{equation}\label{eq3.30}
\hat{C}\hat{O}(\hat{C}^{1/2}\vert\lambda\rangle) =\lambda(\hat{C}^{1/2}\vert\lambda\rangle).
\end{equation}
By $\|\hat{C}^{1/2}\vert\lambda\rangle\|^2=\langle\lambda\vert\hat{C}\vert\lambda\rangle\le\mu_1<+\infty$, the state $\hat{C}^{1/2}\vert\lambda\rangle$ exists and is in $\mathscr{D}(\hat{C}^{-1/2})$ (because $\vert\lambda\rangle\in\mathscr{H}$ by assumption). Thus, $\lambda$ is also an eigenvalue of $\hat{C}\hat{O}$ with eigenstate $\hat{C}^{1/2}\vert\lambda\rangle\in\mathscr{D}(\hat{C}^{-1/2})$, from which it follows that the spectrum of $\hat{M}$ is a subset of the spectrum of the restriction of $\hat{C}\hat{O}$ to $\mathscr{D}(\hat{C}^{-1/2})$.

Conversely, let $\vert\phi\rangle\in\mathscr{D}(\hat{C}^{-1/2})$ be an eigenstate of $\hat{C}\hat{O}$ with eigenvalue $\lambda$. Applying $\hat{C}^{-1/2}$ to both sides of
\begin{equation}\label{eq3.31}
\hat{C}\hat{O}\vert\phi\rangle =\lambda\vert\phi\rangle ,
\end{equation}
one obtains
\begin{equation}\label{eq3.32}
\hat{C}^{1/2}\hat{O}\vert\phi\rangle =\lambda\hat{C}^{-1/2}\vert\phi\rangle .
\end{equation}
From $\vert\phi\rangle\in\mathscr{D}(\hat{C}^{-1/2})$ it follows $\vert\lambda\rangle\stackrel{d}{=}\hat{C}^{-1/2}\vert\phi\rangle\in\mathscr{H}$. Using $\vert\phi\rangle =\hat{C}^{1/2}\vert\lambda\rangle$ to rewrite\ (\ref{eq3.32}) in terms of $\vert\lambda\rangle$ only, one gets
\begin{equation}\label{eq3.33}
\hat{C}^{1/2}\hat{O}\hat{C}^{1/2}\vert\lambda\rangle =\hat{M}\vert\lambda\rangle =\lambda\vert\lambda\rangle ,
\end{equation}
from which one finds that $\lambda$ is also an eigenvalue of $\hat{M}$ with eigenstate $\vert\lambda\rangle$. The spectrum of the restriction of $\hat{C}\hat{O}$ to $\mathscr{D}(\hat{C}^{-1/2})$ is thus a subset of the spectrum of $\hat{M}$, which completes the proof of Proposition\ \ref{prop2}. $\square$
%
%
\section{Applications to large local quadratic forms}\label{sec4}
Write $\mathcal{Q}=\pm\langle\varphi\vert\hat{O}\vert\varphi\rangle$ (with a fixed value of $\pm$). Proposition\ \ref{prop1} means that for large $\mathcal{Q}>0$ and under mild conditions, $\varphi(x)/\|\varphi\|_2$ tends to $\overline{\varphi}_\pm(x)/\|\overline{\varphi}_\pm\|_2$ in $L^2$ and in probability. In shorthand,
\begin{equation}\label{eq4.1}
\frac{\varphi(x)}{\|\varphi\|_2}
\stackrel{L^2,\, p}{\rightarrow}
\frac{\overline{\varphi}_\pm(x)}{\|\overline{\varphi}_\pm\|_2}
\ \ \ \ \ (\mathcal{Q}\rightarrow +\infty).
\end{equation}
In this section, we determine the right-hand side of\ (\ref{eq4.1}) explicitly in two particular cases where $\mathcal{Q}$ is a local quadratic form.
\subsection{An old result revisited}\label{sec4a}
Take $\mathscr{H}=L^2(\Lambda)\otimes\mathbb{C}$ (complex scalar fields) with $\Lambda\ni\lbrace 0\rbrace$ a closed subset of $\mathbb{R}^d$. Consider a zero-mean Gaussian random field $\varphi$ on $\Lambda$ with covariance operator $\hat{C}$ such that $\langle x\vert\hat{C}\vert y\rangle =C(x-y)$, (i.e. $\varphi$ is a homogeneous Gaussian random field). One has the following proposition,

\begin{proposition}\label{prop3}
If $C(x-y)$ is continuous at every $x$ and $y$ in $\Lambda$, then $\mathcal{Q}=\vert\varphi(0)\vert^2$ exists almost surely and
\begin{equation}\label{eq4.2}
\frac{\varphi(x)}{\|\varphi\|_2}
\stackrel{L^2,\, p}{\rightarrow}
{\rm e}^{i\theta}\frac{C(x)}{\|C\|_2}\ \ \ \ \ (\mathcal{Q}\rightarrow +\infty),
\end{equation}
where $\theta$ is a random phase uniformly distributed over $\lbrack 0,2\pi\lbrack$.
\end{proposition}

\bigskip
{\it Proof of Proposition\ \ref{prop3}.} First, we prove that the (normalized) eigenfunctions $\phi_i(x)=\langle x\vert\mu_i\rangle$ are continuous in $\Lambda$. From the continuity of $C(x-y)$ it follows that there exists $A>0$ such that for every $x$ and $y$ in $\Lambda$, $\vert C(x-y)\vert\le A$. By the definition of $\phi_i(x)$ one has, for every $x\in\Lambda$,
\begin{eqnarray}\label{eq4.3}
\vert\phi_i(x)\vert &=&\frac{1}{\mu_i}\left\vert\int_\Lambda C(x-y)\phi_i(y)\, d^dy\right\vert
\le\frac{1}{\mu_i}\int_\Lambda \left\vert C(x-y)\phi_i(y)\right\vert\, d^dy \nonumber \\
&\le&\frac{A}{\mu_i}\int_\Lambda\left\vert\phi_i(y)\right\vert\, d^dy\le
\frac{A\vert\Lambda\vert^{1/2}\|\phi_i\|_2}{\mu_i}=
\frac{A\vert\Lambda\vert^{1/2}}{\mu_i}<+\infty ,
\end{eqnarray}
and for every $x,\, x^\prime\in\Lambda$,
\begin{eqnarray}\label{eq4.4}
\vert\phi_i(x^\prime)-\phi_i(x)\vert &=&\frac{1}{\mu_i}
\left\vert\int_\Lambda \left\lbrack C(x^\prime -y)-C(x-y)\right\rbrack\phi_i(y)\, d^dy\right\vert
\nonumber \\
&\le&\frac{1}{\mu_i}\int_\Lambda \left\vert C(x^\prime -y)-C(x-y)\right\vert
\left\vert\phi_i(y)\right\vert\, d^dy .
\end{eqnarray}
Using\ (\ref{eq4.3}) and $\sup_{x,y\in\Lambda}\vert C(x-y)\vert\le A$ one finds that the integrand on the right-hand side of\ (\ref{eq4.4}) is bounded by the constant $2A^2\vert\Lambda\vert^{1/2}/\mu_i$, and by dominated convergence and the continuity of $C(x-y)$,
\begin{equation}\label{eq4.5}
\lim_{x^\prime\rightarrow x}\vert\phi_i(x^\prime)-\phi_i(x)\vert\le
\frac{1}{\mu_i}\int_\Lambda
\lim_{x^\prime\rightarrow x}\left\vert C(x^\prime -y)-C(x-y)\right\vert
\left\vert\phi_i(y)\right\vert\, d^dy =0.
\end{equation}

We can now prove $\mathscr{D}(\hat{C}^{-1/2})\subset C^0(\Lambda)$, the set of continuous functions in $\Lambda$. Let $\vert\varphi\rangle =\sum_i a_i\vert\mu_i\rangle\in\mathscr{D}(\hat{C}^{-1/2})$. One has $\hat{C}^{-1/2}\vert\varphi\rangle =\sum_i(a_i/\sqrt{\mu_i})\vert\mu_i\rangle$, and $\vert\varphi\rangle\in\mathscr{D}(\hat{C}^{-1/2})$ entails
\begin{equation}\label{eq4.6}
\|\hat{C}^{-1/2}\vert\varphi\rangle\|_2^2=\sum_i\frac{\vert a_i\vert^2}{\mu_i}<+\infty .
\end{equation}
Let $\varphi_N(x)=\sum_{i<N}a_i\phi_i(x)$. Since the $\phi_i(x)$ are continuous in $\Lambda$, the $\varphi_N(x)$ are also continuous in $\Lambda$. From
\begin{eqnarray}\label{eq4.7}
\vert\varphi(x)-\varphi_N(x)\vert&=&\left\vert\sum_{i\ge N}a_i\phi_i(x)\right\vert
\le\sum_{i\ge N}\left\vert a_i\phi_i(x)\right\vert \nonumber \\
&=&\sum_{i\ge N}\sqrt{\mu_i}\vert\phi_i(x)\vert\frac{1}{\sqrt{\mu_i}}\vert a_i\vert \nonumber \\
&\le&\left(\sum_{i\ge N}\mu_i\vert\phi(x)\vert^2\right)^{1/2}
\left(\sum_{i\ge N}\frac{\vert a_i\vert^2}{\mu_i}\right)^{1/2} \nonumber \\
&\le&\left(\sum_{i}\mu_i\vert\phi(x)\vert^2\right)^{1/2}
\left(\sum_{i\ge N}\frac{\vert a_i\vert^2}{\mu_i}\right)^{1/2} \nonumber \\
&=&C(0)^{1/2}\left(\sum_{i\ge N}\frac{\vert a_i\vert^2}{\mu_i}\right)^{1/2}
\le A^{1/2}\left(\sum_{i\ge N}\frac{\vert a_i\vert^2}{\mu_i}\right)^{1/2},
\end{eqnarray}
and\ (\ref{eq4.6}) it follows that $\lim_{N\rightarrow +\infty}\varphi_N(x)=\varphi(x)$ uniformly in $\Lambda$. Thus, $\varphi(x)$ is continuous in $\Lambda$ and $\mathscr{D}(\hat{C}^{-1/2})\subset C^0(\Lambda)$.

Since $\mathscr{D}(\hat{C}^{-1/2})$ is the support of the Gaussian measure defined by the covariance operator $\hat{C}$, $\varphi(x)$ is almost surely continuous in $\Lambda$ and $\mathcal{Q}=\vert\varphi(0)\vert^2$ exists almost surely. For every $\vert\varphi\rangle\in\mathscr{D}(\hat{C}^{-1/2})$ we write $\mathcal{Q}=\langle\varphi\vert\hat{O}\vert\varphi\rangle$ where
\begin{equation}\label{eq4.8}
\hat{O}=\vert 0\rangle\langle 0\vert ,
\end{equation}
is a Hermitian operator on $C^0(\Lambda)^\ast$, the dual of $C^0(\Lambda)$. According to Proposition\ \ref{prop2} and Eq.\ (\ref{eq2.4}), we can determine $\vert\overline{\varphi}\rangle$ by solving the eigenvalue problem $\hat{C}\hat{O}\vert\varphi\rangle =\lambda\vert\varphi\rangle$. In $x$-representation this eigenvalue problem reads
\begin{equation}\label{eq4.9}
C(x)\varphi(0)=\lambda\varphi(x),
\end{equation}
the solution to which is $\lambda_1 =C(0)>0$ with $\varphi_1(x)=\alpha C(x)$ where $\alpha$ is a constant which we will take real and positive without loss of generality, and $\lambda_{n\ge 2}=0$ with $\varphi_{n\ge 2}(x)$ such that $\varphi_{n\ge 2}(0)=0$. Note that $C(0)>0$ follows straightforwardly from $C(0)=\sum_i\mu_i\vert\phi_i(x)\vert^2$ for any $x\in\Lambda$, $\mu_i>0$, and none of the $\phi_i(x)$ being identically zero. By Proposition\ \ref{prop2} one has $\vert\varphi_1\rangle =\hat{C}^{1/2}\vert\lambda_1\rangle$, and $\alpha$ is fixed by the normalization of $\vert\lambda_1\rangle$. Namely, writing $\vert\lambda_1\rangle =\hat{C}^{-1/2}\vert\varphi_1\rangle$ and $\vert\varphi_1\rangle =\alpha\hat{C}\vert 0\rangle$,
\begin{eqnarray}\label{eq4.10}
1&=&\langle\lambda_1\vert\lambda_1\rangle
=\langle\varphi_1\vert\hat{C}^{-1}\vert\varphi_1\rangle \nonumber \\
&=&\alpha^2\langle 0\vert\hat{C}\vert 0\rangle =\alpha^2 C(0),
\end{eqnarray}
which yields $\alpha =C(0)^{-1/2}$. Thus, $\overline{\varphi}(x)$ is given by
\begin{equation}\label{eq4.11}
\overline{\varphi}(x)=\frac{t\, C(x)}{C(0)^{1/2}},
\end{equation}
where $t$ is a complex Gaussian random variable with $\langle t\rangle =\langle t^2\rangle =0$ and  $\langle\vert t\vert^2\rangle =1$.

It is now easy to check that the hypotheses required in Proposition\ \ref{prop1} are fulfilled: ${\rm Tr}\vert\hat{M}\vert ={\rm Tr}\hat{M}=\lambda_1=C(0)<+\infty$ and ${\rm Tr}\vert\hat{C}\vert ={\rm Tr}\hat{C}=\int_\Lambda \langle x\vert\hat{C}\vert x\rangle\, d^dx=\vert\Lambda\vert C(0)<+\infty$. Thus, Proposition\ \ref{prop1} holds and injecting\ (\ref{eq4.11}) into\ (\ref{eq4.1}) one obtains\ (\ref{eq4.2}) where $\theta ={\rm Arg}(t)$ is a random phase uniformly distributed over $\lbrack 0,2\pi\lbrack$, which completes the proof of proposition\ \ref{prop3}. $\square$

\bigskip
The quasi-deterministic behavior of $\varphi$ for large $\mathcal{Q}$ is very clear: the only random element remaining on the right-hand side of\ (\ref{eq4.2}) is the phase $\theta$. In the case of a real scalar field, this result has been known for long (see e.g. Secs. 6.7 and 6.8 in\ \cite{Adl2}), but for a smaller class of smoother fields, with twice derivable correlation functions at $x=0$, and with a stronger, pointwise, convergence. Proposition\ \ref{prop3} enlarges the class of $\varphi$ having this quasi-deterministic behavior. The price to pay is the weaker convergence, in $L^2$, with which the result is to be understood.
\subsection{Gaussian random flow with a large local helicity}\label{sec4b}
Take $\mathscr{H}=L^2(\Lambda)\otimes\mathbb{R}^3$ (real vector fields) with $\Lambda\ni\lbrace 0\rbrace$ a closed convex subset of $\mathbb{R}^3$. We consider a stationary homogeneous Gaussian random flow on $\Lambda$, $\varphi(x)\equiv {\bm v}({\bm x})$, with zero mean and covariance operator $\hat{C}_{\mu\nu}$ such that $\langle {\bm x}\vert\hat{C}_{\mu\nu}\vert{\bm y}\rangle =C_{\mu\nu}({\bm x}-{\bm y})$. In the following, we take
\begin{equation}\label{eq4.13}
C_{\mu\nu}({\bm x})=\frac{2E}{3}\, f(x)\, \delta_{\mu\nu}
+\frac{E}{3}\, xf^\prime (x)\, \left(\delta_{\mu\nu}-\frac{x_\mu x_\nu}{x^2}\right),
\end{equation}
where $x$ denotes $\vert{\bm x}\vert$ from now on, $E$ is the kinetic energy of the turbulent flow per unit mass of fluid, and $f(x)$ is a derivable even function of $x$ with small $x$ behavior
\begin{equation}\label{eq4.14}
f(x)\sim 1-\frac{x^2}{2\ell^2}\ \ \ \ \ (x\rightarrow 0),
\end{equation}
where $\ell$ is the so-called `Taylor microscale', and such that $xf^{\prime\prime\prime}(x)$ exists in a neighborhood of ${\bm x}=0$ with $\lim_{x\rightarrow 0}xf^{\prime\prime\prime}(x)=0$. Expression\ (\ref{eq4.13}) is compatible with the correlation function of an incompressible, isotropic, and homogeneous turbulent flow. Let $h({\bm x})$ denote the local helicity of the flow at point ${\bm x}$ defined by
\begin{equation}\label{eq4.15}
h({\bm x}) ={\bm v}({\bm x})\cdot {\rm curl}\, {\bm v}({\bm x}).
\end{equation}
One has the following proposition

\begin{proposition}\label{prop4}
If $f(\vert {\bm x}-{\bm y}\vert)$ is a $C^2$ function of $\vert {\bm x}-{\bm y}\vert$ at every ${\bm x}$ and ${\bm y}$ in $\Lambda$, then $\mathcal{Q}=\vert h(0)\vert$ exists almost surely and
\begin{equation}\label{eq4.16}
\frac{{\bm v}({\bm x})}{\|{\bm v}\|_2}
\stackrel{L^2,\, p}{\rightarrow}
\frac{{\bm u}_\pm({\bm x})}{\|{\bm u}_\pm\|_2}
\ \ \ \ \ (\mathcal{Q}\rightarrow +\infty),
\end{equation}
with
\begin{eqnarray}\label{eq4.17}
{\bm u}_\pm({\bm x})&=&f(x)\, {\bm e}_t+\frac{x}{2}f^\prime(x)
\left\lbrack {\bm e}_t-({\bm e}_x\cdot{\bm e}_t)\, {\bm e}_x\right\rbrack \\
&\pm&\frac{\ell}{\sqrt{5}}\left\lbrack 2f^\prime(x)+\frac{x}{2}f^{\prime\prime}(x)\right\rbrack
({\bm e}_x\times{\bm e}_t), \nonumber
\end{eqnarray}
where $\pm$ is the sign of $h(0)$, and ${\bm e}_t$ is a random unit vector uniformly distributed over the unit sphere, ${\bm e}_x={\bm x}/x$.
\end{proposition}

\bigskip
\noindent
{\it Remark} : if $\Lambda$ is a ball centered at ${\bm x}=0$ it can easily be checked that $\|{\bm u}_\pm\|_2$ is non random. We leave the proof of this remark to the reader.

\bigskip
{\it Proof of Proposition\ \ref{prop4}.} Note that Eq.\ (\ref{eq4.16}) is not exactly the same as Proposition\ \ref{prop1} written in the form\ (\ref{eq4.1}). In the former, the value of $\pm$ in $\mathcal{Q}=\pm h(0)$ is the sign of $h(0)$ which is {\it not} fixed, while in the latter, the value of $\pm$ is fixed. Let $\sigma$ denote the sign of $h(0)$. The link between Eq.\ (\ref{eq4.16}) and Proposition\ \ref{prop1} is given by the inequality
\begin{eqnarray}\label{link}
\mathbb{P}\left.\left(\left\|\frac{\bm{v}}{\|\bm{v}\|_2}-\frac{\overline{\bm{v}}_\sigma}{\|\overline{\bm{v}}_\sigma\|_2}\right\|_2
>\varepsilon\ \right\vert\ \mathcal{Q}>u\right)&\le&
\mathbb{P}\left.\left(\left\|\frac{\bm{v}}{\|\bm{v}\|_2}-\frac{\overline{\bm{v}}_{+}}{\|\overline{\bm{v}}_{+}\|_2}\right\|_2
>\varepsilon\ \right\vert\ h(0)>u\right) \\
&+&\mathbb{P}\left.\left(\left\|\frac{\bm{v}}{\|\bm{v}\|_2}-\frac{\overline{\bm{v}}_{-}}{\|\overline{\bm{v}}_{-}\|_2}\right\|_2
>\varepsilon\ \right\vert\ -h(0)>u\right), \nonumber
\end{eqnarray}
from which it is readily seen that Proposition\ \ref{prop1} implies Eq.\ (\ref{eq4.16}) with $\bm{u}_\pm =\overline{\bm{v}}_\pm$.

First, we prove that the (normalized) eigenvectors ${\bm v}_i({\bm x})=\langle {\bm x}\vert\mu_i\rangle$ are in $C^1(\Lambda)\otimes\mathbb{R}^3$. Let ${\bm c}_\mu({\bm x})$ denote the three vectors defined by their coordinates $({\bm c}_\mu({\bm x}))_\nu =C_{\mu\nu}({\bm x})$. Since $f(\vert {\bm x}-{\bm y}\vert)$ is a $C^2$ function of $\vert {\bm x}-{\bm y}\vert$ at every ${\bm x}$ and ${\bm y}$ in $\Lambda$, it follows from\ (\ref{eq4.13}) that every component $C_{\mu\nu}({\bm x}-{\bm y})$ is continuous at every ${\bm x}$ and ${\bm y}$ in $\Lambda$. Thus, there exists $A>0$ such that for every index $\mu$, and every ${\bm x}$ and ${\bm y}$ in $\Lambda$, $\vert {\bm c}_\mu({\bm x}-{\bm y})\vert\le A$. By the definition of ${\bm v}_i({\bm x})$ one has, for every index $\mu$ and ${\bm x}\in\Lambda$,
\begin{eqnarray}\label{eq4.18}
\vert ({\bm v}_i({\bm x}))_\mu\vert &=&
\frac{1}{\mu_i}\left\vert\int_\Lambda {\bm c}_\mu({\bm x}-{\bm y})\cdot
{\bm v}_i({\bm y})\, d^3y\right\vert
\le\frac{1}{\mu_i}\int_\Lambda \left\vert {\bm c}_\mu({\bm x}-{\bm y})\vert\, \vert
{\bm v}_i({\bm y})\right\vert\, d^3y \nonumber \\
&\le&\frac{A}{\mu_i}\int_\Lambda\left\vert {\bm v}_i({\bm y})\right\vert\, d^3y\le
\frac{A\vert\Lambda\vert^{1/2}\|{\bm v}_i\|_2}{\mu_i}=
\frac{A\vert\Lambda\vert^{1/2}}{\mu_i}<+\infty .
\end{eqnarray}
Let ${\bm x}$ and ${\bm x}^\prime$ in $\Lambda$ such that ${\bm x}^\prime -{\bm x}$ is along the direction corresponding to the index $\lambda$. One has,
\begin{equation}\label{eq4.19}
\frac{({\bm v}_i({\bm x}^\prime))_\mu-({\bm v}_i({\bm x}))_\mu}
{x^\prime_\lambda -x_\lambda}=\frac{1}{\mu_i}
\int_\Lambda \frac{{\bm c}_\mu ({\bm x}^\prime -{\bm y})-{\bm c}_\mu ({\bm x} -{\bm y})}
{x^\prime_\lambda -x_\lambda}\cdot {\bm v}_i({\bm y})\, d^3y.
\end{equation}
By\ (\ref{eq4.13}),\ (\ref{eq4.14}), and $f(\vert {\bm x}-{\bm y}\vert)$ being a $C^2$ function of $\vert {\bm x}-{\bm y}\vert$ at every ${\bm x}$ and ${\bm y}$ in $\Lambda$, the partial derivatives $\partial C_{\mu\nu}({\bm x}-{\bm y})/\partial x_\lambda$ exist and are continuous at every ${\bm x}$ and ${\bm y}$ in $\Lambda$. It follows in particular that there exists $B>0$ such that for every indices $\mu$ and $\lambda$, and every ${\bm x}$ and ${\bm y}$ in $\Lambda$, $\vert \partial {\bm c}_\mu({\bm x}-{\bm y})/\partial x_\lambda\vert\le B$, which together with $\Lambda$ being convex and the mean value theorem yields the bound
\begin{equation}\label{eq4.20}
\left\vert\frac{{\bm c}_\mu ({\bm x}^\prime -{\bm y})-{\bm c}_\mu ({\bm x} -{\bm y})}
{x^\prime_\lambda -x_\lambda}\right\vert\le B.
\end{equation}
As a result of\ (\ref{eq4.18}) and\ (\ref{eq4.20}), the integrand on the right-hand side of\ (\ref{eq4.19}) is bounded in absolute value by the constant $\sqrt{3}AB\vert\Lambda\vert^{1/2}/\mu_i$ and by dominated convergence applied to\ (\ref{eq4.19}), one finds that $\partial {\bm v}_i({\bm x})/\partial x_\lambda$ exists in $\Lambda$ and is given by
\begin{equation}\label{eq4.21}
\left(\frac{\partial {\bm v}_i({\bm x})}{\partial x_\lambda}\right)_\mu=\frac{1}{\mu_i}
\int_\Lambda \frac{\partial {\bm c}_\mu ({\bm x} -{\bm y})}
{\partial x_\lambda}\cdot {\bm v}_i({\bm y})\, d^3y.
\end{equation}
The continuity of $\partial {\bm v}_i({\bm x})/\partial x_\lambda$ in $\Lambda$ can be proved easily from\ (\ref{eq4.21}) along the same line as the one leading to the continuity of $\varphi_i(x)$ in Sec.\ \ref{sec4a} [Equation\ (\ref{eq4.5})] using\ (\ref{eq4.18}), $\sup_{{\bm x},{\bm y}\in\Lambda ;\mu ,\lambda}\vert \partial {\bm c}_\mu({\bm x}-{\bm y})/\partial x_\lambda\vert\le B$, and $\partial {\bm c}_\mu ({\bm x} -{\bm y})/\partial x_\lambda$ being continuous at every ${\bm x}$ and ${\bm y}$ in $\Lambda$.

We can now prove $\mathscr{D}(\hat{C}^{-1/2})\subset C^1(\Lambda)\otimes\mathbb{R}^3$. Let ${\bm v}({\bm x})=\langle {\bm x}\vert v\rangle$ with $\vert v\rangle =\sum_i a_i\vert\mu_i\rangle\in\mathscr{D}(\hat{C}^{-1/2})$. One has $\hat{C}^{-1/2}\vert v\rangle =\sum_i(a_i/\sqrt{\mu_i})\vert\mu_i\rangle$, and $\vert v\rangle\in\mathscr{D}(\hat{C}^{-1/2})$ entails
\begin{equation}\label{eq4.22}
\|\hat{C}^{-1/2}\vert v\rangle\|_2^2=\sum_i\frac{a_i^2}{\mu_i}<+\infty .
\end{equation}
Let ${\bm v}_N({\bm x})=\sum_{i<N}a_i {\bm v}_i({\bm x})$ (with ${\bm v}_i({\bm x})=\langle {\bm x}\vert\mu_i\rangle$). The counterpart of\ (\ref{eq4.7}) for ${\bm v}({\bm x})$ reads
\begin{equation}\label{eq4.23}
\vert {\bm v}({\bm x})-{\bm v}_N({\bm x})\vert
\le\left(\sum_{\mu}C_{\mu\mu}(0)\right)^{1/2}
\left(\sum_{i\ge N}\frac{a_i^2}{\mu_i}\right)^{1/2}
= (2E)^{1/2}\left(\sum_{i\ge N}\frac{a_i^2}{\mu_i}\right)^{1/2},
\end{equation}
where we have used\ (\ref{eq4.13}) and\ (\ref{eq4.14}) to write $\sum_{\mu}C_{\mu\mu}(0)= 2E$, and from\ (\ref{eq4.22}) it follows that $\lim_{N\rightarrow +\infty}{\bm v}_N({\bm x})={\bm v}({\bm x})$ uniformly in $\Lambda$. Now, for every index $\nu$ consider the sequence $\partial {\bm v}_N({\bm x})/\partial x_\nu=\sum_{i<N}a_i \partial {\bm v}_i({\bm x})/\partial x_\nu$. Since the ${\bm v}_i({\bm x})$ are in $C^1(\Lambda)\otimes\mathbb{R}^3$, the functions of this sequence are continuous in $\Lambda$. Proceeding similarly to\ (\ref{eq4.7}) and using
\begin{equation}\label{eq4.24}
\left.\sum_\mu \frac{\partial^2 C_{\mu\mu}({\bm x}-{\bm y})}
{\partial x_\nu\partial y_\nu}\right\vert_{{\bm y}={\bm x}}=\frac{10E}{3\ell^2},
\end{equation}
which follows from\ (\ref{eq4.13}) and\ (\ref{eq4.14}), one obtains for every integers $M$, $N$, with $M>N$,
\begin{eqnarray}\label{eq4.25}
\left\vert\frac{{\bm v}_M({\bm x})}{\partial x_\nu}-
\frac{{\bm v}_N({\bm x})}{\partial x_\nu}\right\vert
&\le&\left(\left.\sum_\mu \frac{\partial^2 C_{\mu\mu}({\bm x}-{\bm y})}
{\partial x_\nu\partial y_\nu}\right\vert_{{\bm y}={\bm x}}\right)^{1/2}
\left(\sum_{N\le i<M}\frac{a_i^2}{\mu_i}\right)^{1/2} \nonumber \\
&=& \left(\frac{10E}{3\ell^2}\right)^{1/2}\left(\sum_{N\le i<M}\frac{a_i^2}{\mu_i}\right)^{1/2}.
\end{eqnarray}
Finally, by\ (\ref{eq4.25}),\ (\ref{eq4.22}), and $\lim_{N\rightarrow +\infty}{\bm v}_N({\bm x})={\bm v}({\bm x})$, the sequence of continuous functions $\partial {\bm v}_N({\bm x})/\partial x_\nu$ converges uniformly in $\Lambda$ to $\partial {\bm v}({\bm x})/\partial x_\nu$ which is thus continuous in $\Lambda$, i.e. ${\bm v}({\bm x})\in C^1(\Lambda)\otimes\mathbb{R}^3$, hence $\mathscr{D}(\hat{C}^{-1/2})\subset C^1(\Lambda)\otimes\mathbb{R}^3$.

Since $\mathscr{D}(\hat{C}^{-1/2})$ is the support of the Gaussian measure defined by the covariance operator $\hat{C}$, ${\bm v}({\bm x})$ is almost surely in $C^1(\Lambda)\otimes\mathbb{R}^3$ and $\mathcal{Q}=\vert h(0)\vert$ exists almost surely. For every $\vert v\rangle\in\mathscr{D}(\hat{C}^{-1/2})$ we rewrite $\mathcal{Q}$ as $\mathcal{Q}=\vert\langle v\vert\hat{O}^S\vert v\rangle\vert$ where
\begin{equation}\label{eq4.26}
\hat{O}^S=\frac{1}{2}\left(\vert 0\rangle\langle 0\vert\, \hat{{\rm curl}}
+\hat{{\rm curl}}\, \vert 0\rangle\langle 0\vert\right) ,
\end{equation}
is the symmetric part of the operator $\hat{O}=(\vert 0\rangle\langle 0\vert\, \hat{{\rm curl}})$ on $(C^1(\Lambda)\otimes\mathbb{R}^3)^\ast$, the dual of $C^1(\Lambda)\otimes\mathbb{R}^3$. According to Proposition\ \ref{prop2} and Eq.\ (\ref{eq2.4}), we can determine $\vert\overline{v}\rangle$ by solving the eigenvalue problem $\hat{C}\hat{O}^S\vert v\rangle =\lambda\vert v\rangle$. In ${\bm x}$-representation and using the vectors ${\bm c}_\mu({\bm x})$ defined by their coordinates $({\bm c}_\mu({\bm x}))_\nu =C_{\mu\nu}({\bm x})$, this eigenvalue problem reads
\begin{equation}\label{eq4.27}
{\bm c}_\mu({\bm x})\cdot{\rm curl}\, {\bm v}(0)-{\bm v}(0)\cdot{\rm curl}\, {\bm c}_\mu({\bm x})
=2\lambda v_\mu({\bm x}).
\end{equation}
Injecting\ (\ref{eq4.13}) into\ (\ref{eq4.27}), one obtains
\begin{eqnarray}\label{eq4.28}
2\lambda\, {\bm v}({\bm x})&=&\frac{E}{3}\left\lbrack 2f(x)+xf^\prime(x)\right\rbrack {\rm curl}\, {\bm v}(0) \nonumber \\
&-&\frac{E}{3}{\bm x} f^\prime(x)\left(\frac{{\bm x}}{x}\cdot{\rm curl}\, {\bm v}(0)\right) \\
&+&\frac{E}{3}\left\lbrack 4f^\prime(x)+xf^{\prime\prime}(x)\right\rbrack\left(\frac{{\bm x}}{x}\times{\bm v}(0)\right) . \nonumber
\end{eqnarray}
Taking the curl of\ (\ref{eq4.28}) for $x$ small enough, one gets
\begin{eqnarray}\label{eq4.29}
2\lambda\, {\rm curl}{\bm v}({\bm x}) &=&
\frac{E}{3}\left\lbrack 4f^\prime(x)+xf^{\prime\prime}(x)\right\rbrack\left(\frac{{\bm x}}{x}\times{\rm curl}{\bm v}(0)\right) \nonumber \\
&-&\frac{2E}{3}\left\lbrack\frac{4f^\prime(x)}{x}+f^{\prime\prime}(x)\right\rbrack {\bm v}(0) \\
&+&\frac{E}{3}\left\lbrack\frac{4f^\prime(x)}{x}-4f^{\prime\prime}(x)-xf^{\prime\prime\prime}(x)\right\rbrack
\left\lbrack {\bm v}(0) -\left(\frac{{\bm x}}{x}\cdot{\bm v}(0)\right)\frac{{\bm x}}{x}\right\rbrack . \nonumber
\end{eqnarray}
Eqs.\ (\ref{eq4.28}) and\ (\ref{eq4.29}) at ${\bm x}=0$ reduce to the system
\begin{subeqnarray}\label{eq4.30}
&&\lambda\, {\bm v}(0)-\frac{E}{3}\, {\rm curl}{\bm v}(0)=0, \slabel{eq10a} \\
&&\frac{5E}{3\ell^2}\, {\bm v}(0)-\lambda\, {\rm curl}{\bm v}(0) =0, \slabel{eq10b}
\end{subeqnarray}
the solution to which is
\begin{equation}\label{eq4.31}
\lambda_{\pm 1} =\pm \frac{\sqrt{5}\, E}{3\ell},
\end{equation}
with
\begin{equation}\label{eq4.32}
{\rm curl}{\bm v}_{\pm 1}(0)=\pm\frac{\sqrt{5}}{\ell}\, {\bm v}_{\pm 1}(0),
\end{equation}
and $\lambda_{\pm n}=0$ for $n\ge 2$, with ${\rm curl}{\bm v}_{\pm n}(0)={\bm v}_{\pm n}(0)=0$. Note that, according to\ (\ref{eq4.31}) and\ (\ref{eq4.32}), $\lambda_{\pm 1}$ and the helicity of ${\bm v}_{\pm 1}({\bm x})$ at ${\bm x}=0$ have the same sign. From\ (\ref{eq4.27}), (\ref{eq4.31}), and (\ref{eq4.32}) one finds that there are thee different eigenstates with the same eigenvalue $\lambda_{\pm 1}$ given by
\begin{equation}\label{eq4.33}
({\bm v}_{\pm 1}^{(\lambda)}({\bm x}))_\mu =\alpha_\lambda\left\lbrack
{\bm e}_\lambda\cdot {\bm c}_\mu({\bm x})\mp\frac{\ell}{\sqrt{5}}\, {\bm e}_\lambda\cdot{\rm curl}\, {\bm c}_\mu({\bm x})\right\rbrack ,
\end{equation}
where the vectors ${\bm e}_\lambda$ are defined by their coordinates $({\bm e}_\lambda)_\nu =\delta_{\lambda\nu}$, and the $\alpha_\lambda$ are constants. (Here and in the following $\lambda$ is an index not to be confused with the eigenvalue in\ (\ref{eq4.27})). Writing $\vert 0\rangle_\lambda =\vert 0\rangle\otimes {\bm e}_\lambda$ and using the identities
\begin{equation}\label{eq4.34}
\begin{array}{l}
{\bm e}_\lambda\cdot{\bm c}_\mu({\bm x})=
(\langle {\bm x}\vert\hat{C}\vert 0\rangle_\lambda)_\mu , \\
{\bm e}_\lambda\cdot {\rm curl}{\bm c}_\mu({\bm x})=
-(\langle {\bm x}\vert\hat{C}\, \hat{{\rm curl}}\vert 0\rangle_\lambda)_\mu ,
\end{array}
\end{equation}
one can rewrite\ (\ref{eq4.33}) in the representation free form
\begin{equation}\label{eq4.35}
\vert v_{\pm 1}^{(\lambda)}\rangle =\alpha_\lambda
\hat{C}\left( 1\pm\frac{\ell}{\sqrt{5}}\, \hat{{\rm curl}}\right)\vert 0\rangle_\lambda .
\end{equation}
By Proposition\ \ref{prop2} one has $\vert v_{\pm 1}^{(\lambda)}\rangle =\hat{C}^{1/2}\vert\lambda_{\pm 1}^{(\lambda)}\rangle$ and $\alpha_\lambda$ is fixed by the normalization $\langle\lambda_{\pm 1}^{(\lambda)}\vert\lambda_{\pm 1}^{(\lambda)}\rangle =1$. Writing $\vert\lambda_{\pm 1}^{(\lambda)}\rangle =\hat{C}^{-1/2}\vert v_{\pm 1}^{(\lambda)}\rangle$ and using\ (\ref{eq4.35}),\ (\ref{eq4.33}), and\ (\ref{eq4.13}), one finds after some tedious but straightforward algebra,
\begin{eqnarray}\label{eq4.36}
&&\langle\lambda_{\pm 1}^{(\gamma)}\vert\lambda_{\pm 1}^{(\lambda)}\rangle =
(\alpha_\gamma \alpha_\lambda)\, 
_\gamma\langle 0\vert\left( 1\pm\frac{\ell}{\sqrt{5}}\, \hat{{\rm curl}}\right)
\hat{C}\left( 1\pm\frac{\ell}{\sqrt{5}}\, \hat{{\rm curl}}\right)\vert 0\rangle_\lambda \nonumber \\
&&=(\alpha_\gamma \alpha_\lambda)\, 
\left\lbrack {\bm e}_\gamma \cdot{\bm v}_{\pm 1}^{(\lambda)}(0)\pm\frac{\ell}{\sqrt{5}}\, 
{\bm e}_\gamma \cdot{\rm curl}{\bm v}_{\pm 1}^{(\lambda)}(0)\right\rbrack \nonumber \\
&&=(\alpha_\gamma \alpha_\lambda)\, 
\left\lbrack C_{\gamma\lambda}(0)-\frac{\ell^2}{5}\sum_{b,b^\prime ,c,c^\prime}
\varepsilon_{\gamma bc}\varepsilon_{\lambda b^\prime c^\prime}
\frac{\partial^2 C_{cc^\prime}(0)}{\partial x_b \partial x_b^\prime} \right. \nonumber \\
&&\left.\pm\frac{\ell}{\sqrt{5}}\sum_{b,c}
\left(\varepsilon_{\gamma bc}\frac{\partial C_{\lambda c}(0)}{\partial x_b}
-\varepsilon_{\lambda bc}\frac{\partial C_{\gamma c}(0)}{\partial x_b}\right)\right\rbrack
=\frac{4E\alpha_\gamma \alpha_\lambda}{3}\, \delta_{\gamma\lambda},
\end{eqnarray}
where $\varepsilon_{ijk}$ denotes the Levi-Civita symbol, hence $\alpha_\lambda =\sqrt{3}/2\sqrt{E}$. Now, according to\ (\ref{eq2.4}) and\ (\ref{eq4.33}), $\overline{{\bm v}}_\pm({\bm x})$ is given by
\begin{equation}\label{eq4.37}
(\overline{{\bm v}}_\pm({\bm x}))_\mu=
\frac{\sqrt{3}}{2\sqrt{E}}\, \sum_{\lambda =1}^3
t_\lambda\, \left\lbrack
{\bm e}_\lambda\cdot {\bm c}_\mu({\bm x})\mp\frac{\ell}{\sqrt{5}}\, {\bm e}_\lambda\cdot{\rm curl}\, {\bm c}_\mu({\bm x})\right\rbrack ,
\end{equation}
where the $t_\lambda$ are i.i.d. Gaussian random variable with $\langle t_\lambda\rangle =0$ and  $\langle t_\lambda^2\rangle =1$. Injecting\ (\ref{eq4.13}) into\ (\ref{eq4.37}), defining the Gaussian random vector ${\bm t}=(t_1,t_2,t_3)$, and writing ${\bm e}_x={\bm x}/x$ one obtains
\begin{eqnarray}\label{eq4.38}
\overline{{\bm v}}_\pm({\bm x})&=&\sqrt{\frac{2E}{3}}\, \left\lbrace
f(x)\, {\bm t}+\frac{x}{2}f^\prime(x)
\left\lbrack {\bm t}-({\bm e}_x\cdot{\bm t})\, {\bm e}_x\right\rbrack\right. \\
&&\left.\pm\frac{\ell}{\sqrt{5}}\left\lbrack 2f^\prime(x)+\frac{x}{2}f^{\prime\prime}(x)\right\rbrack
({\bm e}_x\times{\bm t})\right\rbrace. \nonumber
\end{eqnarray}

The two hypotheses required in Proposition\ \ref{prop1} are trivially fulfilled: ${\rm Tr}\vert\hat{M}\vert =3\vert\lambda_{\pm 1}\vert=\sqrt{5}E/\ell <+\infty$ and ${\rm Tr}\vert\hat{C}\vert ={\rm Tr}\hat{C}=\sum_{\mu =1}^3\int_\Lambda \langle{\bm x}\vert C_{\mu\mu}\vert{\bm x}\rangle\, d^3x=2\vert\Lambda\vert E<+\infty$. Thus, Proposition\ \ref{prop1} holds and injecting\ (\ref{eq4.38}) into the $u\rightarrow +\infty$ limit of\ (\ref{link}) one obtains\ (\ref{eq4.16}) and\ (\ref{eq4.17}). Finally, since ${\bm t}$ is the Gaussian random vector ${\bm t}=(t_1,t_2,t_3)$, ${\bm e}_t={\bm t}/\vert {\bm t}\vert$ is a random unit vector uniformly distributed over the unit sphere, which completes the proof of proposition\ \ref{prop4}. $\square$
%
%
\section{Summary}\label{sec5}
In this paper, we have studied the realizations of a Gaussian random field, $\varphi$, in the limit where some real quadratic form $\langle\varphi\vert\hat{O}\vert\varphi\rangle$ is large. More specifically, for $\vert\langle\varphi\vert\hat{O}\vert\varphi\rangle\vert\rightarrow +\infty$ and under mild conditions, we have proved concentration in $L^2$ and in probability of $\varphi/\|\varphi\|_2$ along the components associated with the largest eigenvalue $\lambda_{\pm 1}$ of the restriction of $\pm\hat{C}\hat{O}$ to the domain of $\hat{C}^{-1/2}$ (Propositions\ \ref{prop1} and\ \ref{prop2}). Here, $\pm$ is the sign of $\langle\varphi\vert\hat{O}\vert\varphi\rangle$ and $\hat{C}$ is the correlation operator of $\varphi$. If the degeneracy of $\lambda_{\pm 1}$ is low, this concentration results in a quasi-deterministic behavior of $\varphi$. Propositions\ \ref{prop1} and\ \ref{prop2} generalize the proposition 1 proved by MC in\ \cite{MC} for $\hat{O}=\hat{1}$.

We have then used these results to determine the quasi-deterministic behavior of $\varphi$ explicitly in two particular cases where the quadratic form $\langle\varphi\vert\hat{O}\vert\varphi\rangle$ is local. First, we have considered a homogeneous complex scalar field under the condition that $\vert\varphi(0)\vert^2$ is large. In this limit, we have found that $\varphi$ becomes simply proportional to its correlation function, to within a random phase factor which is the only remaining random element (Proposition\ \ref{prop3}). In the case of a real $\varphi$, this result has been known for long (see e.g. Secs. 6.7 and 6.8 in\ \cite{Adl2}), but for a much smaller class of smoother fields, with twice derivable correlation functions at $x=0$, and with a stronger, pointwise, convergence. By requiring a weaker convergence, in $L^2$, Proposition\ \ref{prop3} considerably enlarges the class of $\varphi$ having this quasi-deterministic behavior. Then, we have determined the quasi-deterministic structure of a three-dimensional Gaussian random flow, ${\bm v}({\bm x})$, when its local helicity $h(0)={\bm v}(0)\cdot{\rm curl}{\bm v}(0)$ has a large absolute value (Proposition\ \ref{prop4}). Such a structure can be expected to be of interest e.g. in the context of turbulent dynamo\ \cite{Mof} in any Kraichnan like model of advection, where helicity is a key quantity and the turbulent flow is approximated by a Gaussian random flow. As far as we know, this result (Proposition\ \ref{prop4}) has never been obtained before.

To conclude, note that in view of the importance of Gaussian processes, and more generally Gaussian fields, in statistical modeling, this work is expected to be of interest in a wide range of physical situations. Beside the problem of turbulent flows with a large helicity, it is worth mentioning possible applications to e.g. random media in which a large value of stress or energy can cause damage. An example is the problem of damage in optical materials for high-intensity smoothed laser beams. (Here, the random medium is the pair glass/random laser field). Assuming that the medium can be properly modeled by Gaussian random fields and since stress and energy are quadratic quantities, it follows from our results that damage turns out to be associated with the occurrence of quasi-deterministic structures within the medium. Knowing these structures could help to improve the resistance of the medium with respect to the considered damage. Such applications will be the subject of a future work.
\section*{Acknowledgements} The author warmly thanks Harvey A. Rose for providing valuable insights, and in particular for pointing out the problem solved in Sec.\ \ref{sec4b}, which was one of the motivations for this work. He also thanks Pierre Collet and Satya N. Majumdar for useful discussions, as well as Joel L. Lebowitz for interest and useful discussions on related subjects. Girish Sharma participated in this work as part of his Master's internship in the Erasmus Mundus program under the author's supervision.
%
%
%
%
\appendix*
\section{Proof of proposition\ \ref{prop1} for a real field ($\bm{\mathscr{H}=L^2(\Lambda)\otimes\mathbb{R}^N}$)}\label{app1}
For a real field, i.e. for $\mathscr{H}=L^2(\Lambda)\otimes\mathbb{R}^N$, one has $\langle\varphi\vert\hat{O}\vert\varphi\rangle = \langle\varphi\vert\hat{O}^S\vert\varphi\rangle$ and the symmetric operator to be considered is actually $\hat{O}^S$, the symmetric part of $\hat{O}$. (It follows in particular that $\hat{O}$ needs not be symmetric as it always appears through its symmetric part only).

We give the proof for ``$\pm =+$". (The proof for ``$\pm =-$" is similar). To estimate the two conditional probabilities on the right-hand side of\ (\ref{eq3.2}) we first need to bound $\mathbb{P}(\langle\varphi\vert\hat{O}^S\vert\varphi\rangle >u)$ from below. Let $\rho(v)$ denote the probability distribution function (pdf) of $\sum_{\lbrace 1\le n\le g_1\rbrace\cup\lbrace n<0\rbrace}\lambda_n t_n^2$. One has the integral representation
\begin{equation}\label{ap1.1}
\rho(v)=\int_{-\infty}^{+\infty}\frac{\exp(-ikv)}{(1-2ik\lambda_1)^{g_1/2}}\, \prod_{n>0}\frac{1}{\sqrt{1-2ik\lambda_{-n}}}\, \frac{dk}{2\pi}.
\end{equation}
For $v\rightarrow +\infty$, the leading asymptotic behavior of \ (\ref{ap1.1}) is determined by the vicinity of the singularity at $k=-i/2\lambda_1$ and one gets
\begin{eqnarray}\label{ap1.2}
\rho(v)&\sim&\prod_{n>0}\frac{1}{\sqrt{1-\lambda_{-n}/\lambda_1}}\, 
\int_{-\infty}^{+\infty}\frac{\exp(-ikv)}{(1-2ik\lambda_1)^{g_1/2}}\, \frac{dk}{2\pi} \\
&=&\frac{{\rm e}^{-v/2\lambda_1}}{2\lambda_1}\prod_{n>0}\frac{1}{\sqrt{1-\lambda_{-n}/\lambda_1}}\, 
\int_{1-i\infty}^{1+i\infty}\frac{\exp(vs/2\lambda_1)}{s^{g_1/2}}\, \frac{ds}{2i\pi}\ \ \ \ (v\rightarrow +\infty), \nonumber
\end{eqnarray}
where $s=1-2ik\lambda_1$. (Note that $\hat{M}$ being trace class ensures that the product in front of the integral exist). The remaining $s$-integral is easily obtained from\ \cite{Abra}. One finds
\begin{equation}\label{ap1.3}
\rho(v)\sim\frac{1}{(g_1/2-1)!}\prod_{n>0}\frac{1}{\sqrt{1-\lambda_{-n}/\lambda_1}}\, \left(\frac{v}{2\lambda_1}\right)^{g_1/2-1}
\frac{\exp(-v/2\lambda_1)}{2\lambda_1}\ \ \ \ (v\rightarrow +\infty),
\end{equation}
from which it follows that for every $0<\alpha <1$, there is $v_0 >0$ such that for every $v>v_0$
\begin{equation}\label{ap1.4}
\rho(v)\ge\frac{(1-\alpha)}{(g_1/2-1)!}\prod_{n>0}\frac{1}{\sqrt{1-\lambda_{-n}/\lambda_1}}\, \left(\frac{v}{2\lambda_1}\right)^{g_1/2-1}
\frac{\exp(-v/2\lambda_1)}{2\lambda_1}.
\end{equation}
From\ (\ref{ap1.4}) and the lower bound
\begin{eqnarray}\label{ap1.5}
\mathbb{P}\left(\langle\varphi\vert\hat{O}^S\vert\varphi\rangle >u\right)&\equiv&
\mathbb{P}\left(\sum_n \lambda_n t_n^2 >u\right) \nonumber \\
&\ge&\mathbb{P}\left(\sum_{\lbrace 1\le n\le g_1\rbrace\cup\lbrace n<0\rbrace}\lambda_n t_n^2 >u\right) \nonumber \\
&=&\int_u^{+\infty}\rho(v)\, dv,
\end{eqnarray}
one finds that for $u>v_0$,
\begin{equation}\label{ap1.6a}
\mathbb{P}\left(\langle\varphi\vert\hat{O}^S\vert\varphi\rangle >u\right)\ge \frac{C_1(\alpha)}{1-\alpha}\, 
\int_u^{+\infty}v^{g_1/2-1}
\exp\left(-\frac{v}{2\lambda_1}\right)\, \frac{dv}{2\lambda_1},
\end{equation}
with
\begin{equation}\label{ap1.7}
C_1(\alpha) = \frac{(1-\alpha)^2}{(g_1/2-1)!\, (2\lambda_1)^{g_1/2-1}}
\prod_{n>0}\frac{1}{\sqrt{1-\lambda_{-n}/\lambda_1}}.
\end{equation}
From the asymptotics
\begin{equation*}
\int_u^{+\infty}v^{g_1/2-1}\exp\left(-\frac{v}{2\lambda_1}\right)\, \frac{dv}{2\lambda_1}\sim
u^{g_1/2-1}\exp\left(-\frac{u}{2\lambda_1}\right)\ \ \ \ (u\rightarrow +\infty),
\end{equation*}
it follows that there is $v_1>0$ such that for every $u>v_1$
\begin{equation}\label{ap1.6b}
\int_u^{+\infty}v^{g_1/2-1}\exp\left(-\frac{v}{2\lambda_1}\right)\, \frac{dv}{2\lambda_1}\ge
(1-\alpha)\, u^{g_1/2-1}\exp\left(-\frac{u}{2\lambda_1}\right).
\end{equation}
Injecting\ (\ref{ap1.6b}) into\ (\ref{ap1.6a}), one finds that for $u$ large enough (i.e. $u>\max\lbrace v_0,v_1\rbrace$)
\begin{equation}\label{ap1.6}
\mathbb{P}\left(\langle\varphi\vert\hat{O}^S\vert\varphi\rangle >u\right)\ge C_1(\alpha)\, 
u^{g_1/2-1}\exp\left(-\frac{u}{2\lambda_1}\right).
\end{equation}
[Note that if $g_1>1$, one can as well follow the same calculation as in the complex case, i.e. without using\ (\ref{ap1.6b}), and $(1-\alpha)^2$ is replaced with $1-\alpha$ in\ (\ref{ap1.7})]. First, we estimate the conditional probabilities $\mathbb{P}_u\left(\|\overline{\varphi}\|_2^2<a\right)$. One has
\begin{eqnarray}\label{ap1.8}
&&\mathbb{P}\left(\|\overline{\varphi}\|_2^2 <a,\, \langle\varphi\vert\hat{O}^S\vert\varphi\rangle >u\right) =
\mathbb{P}\left(\sum_{i,j=1}^{g_1}\langle\lambda_i\vert\hat{C}\vert\lambda_j\rangle t_i t_j <a,\, 
\sum_i\lambda_i  t_i^2 >u\right) \nonumber \\
&&\le\mathbb{P}\left(\sum_{i,j=1}^{g_1}\langle\lambda_i\vert\hat{C}\vert\lambda_j\rangle t_i t_j <a,\, 
\sum_{i\ge 1}\lambda_i  t_i^2 >u\right).
\end{eqnarray}
The matrix $\langle\lambda_i\vert\hat{C}\vert\lambda_j\rangle$ is a $g_1\times g_1$ positive definite symmetric (real) matrix. Let $\tilde{\mu}_1\ge\tilde{\mu}_2\ge\cdots\ge\tilde{\mu}_{g_1}$ denote its eigenvalues and $\lbrace\vert\tilde{\mu}_i\rangle\rbrace$ the corresponding orthonormal basis of eigenvectors. For every realization of the $t_i$ one has
\begin{equation}\label{ap1.9}
\sum_{i,j=1}^{g_1}\langle\lambda_i\vert\hat{C}\vert\lambda_j\rangle t_i t_j =\sum_{n=1}^{g_1}\tilde{\mu}_n \tilde{t}_n^2,
\end{equation}
where
\begin{equation}\label{ap1.10}
\tilde{t}_n=\sum_{i=1}^{g_1}\langle\tilde{\mu}_n\vert\lambda_i\rangle t_i.
\end{equation}
From\ (\ref{ap1.10}) one finds that the $\tilde{t}_i$ have the same statistical properties as the $t_i$ (i.e. they are i.i.d. real Gaussian random variables with $\langle\tilde{t}_i\rangle =0$ and  $\langle\tilde{t}_i^2\rangle =1$), with
\begin{equation}\label{ap1.11}
\sum_{i=1}^{g_1} t_i^2 =\sum_{i=1}^{g_1}\tilde{t}_i^2.
\end{equation}
Using\ (\ref{ap1.10}) and\ (\ref{ap1.11}) on the right-hand side of\ (\ref{ap1.8}) and dropping the tilde (because the $\tilde{t}_i$ and the $t_i$ have the same statistical properties), one obtains
\begin{equation}\label{ap1.12}
\mathbb{P}\left(\|\overline{\varphi}\|_2^2 <a,\, \langle\varphi\vert\hat{O}^S\vert\varphi\rangle >u\right)\le
\mathbb{P}\left(\sum_{i=1}^{g_1}\tilde{\mu}_i t_i^2 <a,\, \sum_{i\ge 1}\lambda_i t_i^2 >u\right).
\end{equation}
It follows from $\mu_n>0$ for every $n$ and\ (\ref{eq2.2}) that $\tilde{\mu}_i =\langle\tilde{\mu}_i\vert\hat{C}\vert\tilde{\mu}_i\rangle >0$ for every $i$, and from $\hat{M}$ being trace class that $g_1<+\infty$. Thus,
\begin{equation}\label{ap1.13}
\tilde{\mu}_{min}=\inf_{1\le i\le g_1}\left\lbrace\tilde{\mu}_i\, :\, \tilde{\mu}_i>0\right\rbrace =\tilde{\mu}_{g_1}>0,
\end{equation}
and\ (\ref{ap1.12}) is bounded above by
\begin{eqnarray}\label{ap1.14}
&&\mathbb{P}\left(\|\overline{\varphi}\|_2^2 <a,\, \langle\varphi\vert\hat{O}^S\vert\varphi\rangle >u\right)\le
\mathbb{P}\left(\tilde{\mu}_{min}\sum_{i=1}^{g_1} t_i^2 <a,\, \sum_{i\ge 1}\lambda_i t_i^2 >u\right) \nonumber \\
&&=\int_{x=0}^{+\infty}\mathbb{P}\left(\frac{u-x}{\lambda_1} <\sum_{i=1}^{g_1} t_i^2 <\frac{a}{\tilde{\mu}_{min}}\right)\, 
d\mathbb{P}\left(\sum_{i>g_1}\lambda_i t_i^2 =x\right) \nonumber \\
&&=\int_{x=u-\lambda_1 a/\tilde{\mu}_{min}}^{+\infty}\mathbb{P}\left(\frac{u-x}{\lambda_1} <\sum_{i=1}^{g_1} t_i^2 <\frac{a}{\tilde{\mu}_{min}}\right)\, d\mathbb{P}\left(\sum_{i>g_1}\lambda_i t_i^2 =x\right) \nonumber \\
&&\le\int_{x=u-\lambda_1 a/\tilde{\mu}_{min}}^{+\infty}d\mathbb{P}\left(\sum_{i>g_1}\lambda_i t_i^2 =x\right) \nonumber \\
&&=\mathbb{P}\left(\sum_{i>g_1}\lambda_i t_i^2 >u-\frac{\lambda_1 a}{\tilde{\mu}_{min}}\right),
\end{eqnarray}
where we have used the statistical independence of the $t_i$ (second line), and the fact that the probability in the integrand vanishes identically for $x<u-\lambda_1 a/\tilde{\mu}_{min}$ (third line). Now, by exponential Markov inequality, one has for every positive $c<1/2\lambda_{g_1+1}$,
\begin{eqnarray*}
\mathbb{P}\left(\sum_{i>g_1}\lambda_i t_i^2 >u-\frac{\lambda_1 a}{\tilde{\mu}_{min}}\right)&\le&
{\rm e}^{-c(u-\lambda_1 a/\tilde{\mu}_{min})}\mathbb{E}\left\lbrack\exp\left(c\sum_{i>g_1}\lambda_i t_i^2\right)\right\rbrack \nonumber \\
&=&{\rm e}^{-c(u-\lambda_1 a/\tilde{\mu}_{min})}\prod_{i>g_1}\frac{1}{\sqrt{1-2c\lambda_i}},
\end{eqnarray*}
and by taking $c=(\lambda_1^{-1}+\lambda_{g_1+1}^{-1})/4$, one gets
\begin{equation}\label{ap1.15}
\mathbb{P}\left(\|\overline{\varphi}\|_2^2 <a,\, \langle\varphi\vert\hat{O}^S\vert\varphi\rangle >u\right)\le
C_2(a) \exp\left\lbrack -\left(\frac{1}{\lambda_1}+\frac{1}{\lambda_{g_1+1}}\right)\frac{u}{4}\right\rbrack ,
\end{equation}
with
\begin{equation}\label{ap1.16}
C_2(a)=\exp\left\lbrack\left(1+\frac{\lambda_1}{\lambda_{g_1+1}}\right)\frac{a}{4\tilde{\mu}_{min}}\right\rbrack
\prod_{i>g_1}\frac{1}{\sqrt{1-(\lambda_1^{-1}+\lambda_{g_1+1}^{-1})\lambda_i/2}}.
\end{equation}
The existence of the product on the right-hand side of\ (\ref{ap1.16}) is ensured by $\hat{M}$ being trace class. Finally, it follows from the two estimates\ (\ref{ap1.6}) and\ (\ref{ap1.15}) that for $u$ large enough $\mathbb{P}_u\left(\|\overline{\varphi}\|_2^2<a\right)$ is bounded above by
\begin{eqnarray}\label{ap1.17}
\mathbb{P}_u\left(\|\overline{\varphi}\|_2^2<a\right) &\equiv&
\frac{\mathbb{P}\left(\|\overline{\varphi}\|_2^2 <a,\, \langle\varphi\vert\hat{O}^S\vert\varphi\rangle >u\right)}
{\mathbb{P}\left(\langle\varphi\vert\hat{O}^S\vert\varphi\rangle >u\right)} \nonumber \\
&\le&\frac{C_2(a)}{C_1(\alpha)}\, \frac{1}{u^{g_1/2-1}}\, 
\exp\left\lbrack -\left(\frac{1}{\lambda_{g_1+1}}-\frac{1}{\lambda_1}\right)\frac{u}{4}\right\rbrack .
\end{eqnarray}
We now estimate the limit $\lim_{u\rightarrow +\infty}\mathbb{P}_u\left(\|\delta\varphi\|_2^2>\varepsilon a\right)$. One has
\begin{eqnarray}\label{ap1.18}
&&d\mathbb{P}\left(\lbrace t_{i\notin\lbrack 1,g_1\rbrack}\rbrace ,\, \sum_i \lambda_i t_i^2 >u\right)
\le d\mathbb{P}\left(\lbrace t_{i\notin\lbrack 1,g_1\rbrack}\rbrace ,\, \sum_{i\ge 1} \lambda_i t_i^2 >u\right) \nonumber \\
&&=\mathbb{P}\left(\sum_{i=1}^{g_1} t_i^2 >\frac{u}{\lambda_1}-\sum_{i>g_1}\frac{\lambda_i}{\lambda_1} t_i^2\right)
\, \prod_{i\notin\lbrack 1,g_1\rbrack}\frac{{\rm e}^{-t_i^2/2}}{\sqrt{2\pi}}\, dt_i,
\end{eqnarray}
with
\begin{eqnarray}\label{ap1.19}
&&\mathbb{P}\left(\sum_{i=1}^{g_1} t_i^2 >\frac{u}{\lambda_1}-\sum_{i>g_1}\frac{\lambda_i}{\lambda_1} t_i^2\right) \nonumber \\
&&=\frac{S_{g_1}}{2(2\pi)^{g_1/2}}
\int_{(u-\sum_{i>g_1}\lambda_i t_i^2)/\lambda_1}^{+\infty}H(v)\, v^{g_1/2-1}\, {\rm e}^{-v/2}\, dv,
\end{eqnarray}
where $H(v)$ is the Heaviside step function and $S_{g_1}$ is the unit sphere surface area in a $g_1$-dimensional space ($S_1=2$).

If $g_1>1$ one can follow exactly the same line as in the complex field case. One finds that for every $\alpha >0$, there is $v_2 >0$ such that for every $u>v_2$
\begin{equation}\label{ap1.20}
\mathbb{P}\left(\sum_{i=1}^{g_1} t_i^2 >\frac{u}{\lambda_1}-\sum_{i>g_1}\frac{\lambda_i}{\lambda_1} t_i^2\right)\le
\frac{(1+\alpha)\, S_{g_1}}{(2\pi)^{g_1/2}}\left(\frac{u}{\lambda_1}\right)^{g_1/2-1}\, {\rm e}^{-u/2\lambda_1}
\, \prod_{i>g_1}{\rm e}^{\lambda_i t_i^2/2\lambda_1}.
\end{equation}
Taking $0<\alpha <1$, it follows from\ (\ref{ap1.6}),\ (\ref{ap1.18}), and\ (\ref{ap1.20}) that for $u$ large enough (i.e. $u>\max\lbrace v_0,v_1,v_2\rbrace$)
\begin{equation}\label{ap1.21}
d\mathbb{P}_u\left(\lbrace t_{i\notin\lbrack 1,g_1\rbrack}\rbrace\right)\le C_3(\alpha)
\, \left(\prod_{i<0}\frac{{\rm e}^{-t_i^2/2}}{\sqrt{2\pi}}\, dt_i\right)
\, \left(\prod_{i>g_1}\frac{{\rm e}^{-(1-\lambda_i/\lambda_1)t_i^2/2}}{\sqrt{2\pi}}\, dt_i\right),
\end{equation}
with
\begin{equation}\label{ap1.22}
C_3(\alpha)=\frac{(1+\alpha)\, S_{g_1}}{(2\pi)^{g_1/2}C_1(\alpha)\lambda_1^{g_1/2-1}}.
\end{equation}
Using\ (\ref{ap1.21}) to estimate $\mathbb{E}_u\left\lbrack\exp\left( c\|\delta\varphi\|_2^2\right)\right\rbrack$, one finds that for $u$ large enough
\begin{equation}\label{ap1.23}
\mathbb{E}_u\left\lbrack\exp\left( c\|\delta\varphi\|_2^2\right)\right\rbrack\le C_3(\alpha)C_4(c),
\end{equation}
where $C_4(c)$ is an infinite product of Gaussian integrals the existence of which is ensured if the matrix
\begin{equation*}
{\rm diag}\left(\min\left\lbrace 1,\, 1-\lambda_i/\lambda_1\right\rbrace\right)
-2c\, \langle\lambda_i\vert\hat{C}\vert\lambda_j\rangle ,
\end{equation*}
is (strictly) positive definite, and the matrix
\begin{equation*}
{\rm diag}\left(\max\left\lbrace 0,\, \lambda_i/\lambda_1\right\rbrace\right)
+2c\, \langle\lambda_i\vert\hat{C}\vert\lambda_j\rangle ,
\end{equation*}
is trace class, with $i$ and $j$ $\notin\lbrack 1,\, g_1\rbrack$. The latter condition is fulfilled by $\hat{M}$ and $\hat{C}$ being trace class. The former one requires $0<c<(1-\lambda_{g_1+1}/\lambda_1)/2\mu_1$, which follows from $\|\hat{C}\| = \mu_1 <+\infty$ and $\min\left\lbrace 1,\, 1-\lambda_i/\lambda_1\right\rbrace\ge 1-\lambda_{g_1+1}/\lambda_1>0$ for $i\notin\lbrack 1,\, g_1\rbrack$. Thus, there exists $c>0$ such that $C_4(c)<+\infty$, and by exponential Markov inequality and Eq.\ (\ref{ap1.23}) one finds that for $u$ large enough
\begin{eqnarray}\label{ap1.24}
\mathbb{P}_u\left(\|\delta\varphi\|_2^2>\varepsilon a\right)&\le& {\rm e}^{-\varepsilon ca}
\mathbb{E}_u\left\lbrack\exp\left( c\|\delta\varphi\|_2^2\right)\right\rbrack \nonumber \\
&\le&C_3(\alpha)C_4(c)\, {\rm e}^{-\varepsilon ca}.
\end{eqnarray}
The end of the proof is the same as in the complex case.

If $g_1=1$ a different approach is needed because, in this case, $1/\sqrt{v}$ is not an increasing function of $v$ and one cannot use the arguments leading from\ (\ref{ap1.19}) to\ (\ref{ap1.20}). In the following we write $\sigma =\sum_{i>1}\lambda_i t_i^2$ and for fixed $0<\varepsilon <1$ we consider the two complementary domains $\sigma <(1-\varepsilon)u$ and $\sigma\ge (1-\varepsilon)u$ respectively.

In the first domain, one can bound\ (\ref{ap1.19}) above by making the change of variable $v\rightarrow v-\sigma/\lambda_1$ on the right-hand side of (\ref{ap1.19}) and using the fact that $1/\sqrt{v}$ is a decreasing function of $v$. One obtains
\begin{equation}\label{ap1.25}
\mathbb{P}\left( t_1^2 >\frac{u-\sigma}{\lambda_1}\right)\le
2\sqrt{\frac{\lambda_1}{\varepsilon}}\, \frac{1}{\sqrt{u}}\, {\rm e}^{-u/2\lambda_1}
\, \prod_{i>g_1}{\rm e}^{\lambda_i t_i^2/2\lambda_1},
\end{equation}
which is very similar to\ (\ref{ap1.20}) with a different constant. Taking $0<\alpha <1$, it follows from\ (\ref{ap1.6}),\ (\ref{ap1.18}), and\ (\ref{ap1.25}) that for $u$ large enough (i.e. $u>\max\lbrace v_0,v_1\rbrace$)
\begin{equation}\label{ap1.26}
d\mathbb{P}_u\left(\lbrace t_{i\notin\lbrack 1,g_1\rbrack}\rbrace\right)\le C_3(\alpha)
\, \left(\prod_{i<0}\frac{{\rm e}^{-t_i^2/2}}{\sqrt{2\pi}}\, dt_i\right)
\, \left(\prod_{i>g_1}\frac{{\rm e}^{-(1-\lambda_i/\lambda_1)t_i^2/2}}{\sqrt{2\pi}}\, dt_i\right),
\end{equation}
with
\begin{equation}\label{ap1.27}
C_3(\alpha)=\frac{2}{C_1(\alpha)}\, \sqrt{\frac{\lambda_1}{\varepsilon}}.
\end{equation}
Using first\ (\ref{ap1.26}) and then ${\bm 1}_{\lbrace\sigma<(1-\varepsilon)u\rbrace}\le 1$ to estimate $\mathbb{E}_u\left\lbrack\exp\left( c\|\delta\varphi\|_2^2\right){\bm 1}_{\lbrace\sigma<(1-\varepsilon)u\rbrace}\right\rbrack$, one finds
\begin{equation}\label{ap1.28}
\mathbb{E}_u\left\lbrack\exp\left( c\|\delta\varphi\|_2^2\right){\bm 1}_{\lbrace\sigma<(1-\varepsilon)u\rbrace}\right\rbrack\le C_3(\alpha)C_4(c),
\end{equation}
where $C_4(c)$ is the same infinite product of Gaussian integrals as in\ (\ref{ap1.23}).

In the second domain, we use $\mathbb{P}[t_1^2>(u-\sigma)/\lambda_1]\le 1$ to bound\ (\ref{ap1.18}) trivially by
\begin{equation}\label{ap1.29}
d\mathbb{P}\left(\lbrace t_{i\ne 1}\rbrace ,\, \sum_i \lambda_i t_i^2 >u\right)
\le\prod_{i\ne 1}\frac{{\rm e}^{-t_i^2/2}}{\sqrt{2\pi}}\, dt_i.
\end{equation}
From\ (\ref{ap1.6}),\ (\ref{ap1.29}), and H\"older's inequality it follows that for $u$ large enough,
\begin{eqnarray}\label{ap1.30}
&&\mathbb{E}_u\left\lbrack\exp\left( c\|\delta\varphi\|_2^2\right){\bm 1}_{\lbrace\sigma\ge (1-\varepsilon)u\rbrace}\right\rbrack
\le\frac{\sqrt{u}{\rm e}^{u/2\lambda_1}}{C_1(\alpha)}
\mathbb{E}\left\lbrack\exp\left( c\|\delta\varphi\|_2^2\right){\bm 1}_{\lbrace\sigma\ge (1-\varepsilon)u\rbrace}\right\rbrack \nonumber \\
&&\le\frac{\sqrt{u}{\rm e}^{u/2\lambda_1}}{C_1(\alpha)}
\mathbb{E}\left\lbrack\exp\left( qc\|\delta\varphi\|_2^2\right)\right\rbrack^{1/q}
\mathbb{P}\left\lbrack\sigma\ge (1-\varepsilon)u\right\rbrack^{1/p} \nonumber \\
&&\le\frac{C_4(qc)^{1/q}}{C_1(\alpha)}\sqrt{u}{\rm e}^{u/2\lambda_1}
\mathbb{P}\left\lbrack\sigma\ge (1-\varepsilon)u\right\rbrack^{1/p},
\end{eqnarray}
with $p>1$, $1/p+1/q=1$, and where we have used $\mathbb{E}\left\lbrack\exp\left( c\|\delta\varphi\|_2^2\right)\right\rbrack\le C_4(c)$ (to avoid introducing another constant). The probability on the right-hand side of\ (\ref{ap1.30}) can be estimated from the asymptotic evaluation of its integral representation for large $u$. Skipping the details, one finds that for every $\alpha >0$, there is $v_2>0$ such that for every $u>v_2$
\begin{equation}\label{ap1.31}
\mathbb{P}\left\lbrack\sigma\ge (1-\varepsilon)u\right\rbrack\le (1+\alpha)C_5(\alpha)u^{g_2/2-1}
\exp\left\lbrack -\frac{(1-\varepsilon)u}{2\lambda_2}\right\rbrack,
\end{equation}
where $C_5(\alpha)$ is a constant. Injecting\ (\ref{ap1.31}) into\ (\ref{ap1.30}) one obtains, for $u$ large enough (i.e. $u>\max\lbrace v_0,v_1,v_2\rbrace$)
\begin{eqnarray}\label{ap1.32}
&&\mathbb{E}_u\left\lbrack\exp\left( c\|\delta\varphi\|_2^2\right){\bm 1}_{\lbrace\sigma\ge (1-\varepsilon)u\rbrace}\right\rbrack
\le (1+\alpha)\, \frac{C_5(\alpha)C_4(qc)^{1/q}}{C_1(\alpha)} \nonumber \\
&&\times u^{1/2+g_2/2p-1/p}\exp\left\lbrack -\left(\frac{1-\varepsilon}{p\lambda_2}-\frac{1}{\lambda_1}\right)\, \frac{u}{2}\right\rbrack .
\end{eqnarray}
Write
\begin{equation*}
\gamma(\varepsilon ,p)=\frac{1}{2}\, \left(\frac{1-\varepsilon}{p\lambda_2}-\frac{1}{\lambda_1}\right).
\end{equation*}
It remains to fix $0<\varepsilon <1-\lambda_2/\lambda_1$, $1<p<(1-\varepsilon)\lambda_1/\lambda_2$, and $0<c<(1-\lambda_{g_1+1}/\lambda_1)/2q\mu_1$ and one has, for $u$ large enough,
\begin{equation}\label{ap1.33}
\mathbb{E}_u\left\lbrack\exp\left( c\|\delta\varphi\|_2^2\right){\bm 1}_{\lbrace\sigma\ge (1-\varepsilon)u\rbrace}\right\rbrack
\le C_6(\alpha)\, u^{1/2+g_2/2p-1/p} {\rm e}^{-\gamma(\varepsilon ,p)u},
\end{equation}
with $\gamma(\varepsilon ,p)>0$ and where $C_6(\alpha)$ is a constant. Note that $c$ is chosen such that $C_4(qc)$ on the right-hand side of\ (\ref{ap1.32}) exists, and since $C_4(c)\le C_4(qc)$ by $q>1$, it follows that $C_4(c)$ on the right-hand side of\ (\ref{ap1.28}) exists too. By exponential Markov inequality, Eq.\ (\ref{ap1.28}), and Eq.\ (\ref{ap1.33}) one finds that for $u$ large enough
\begin{eqnarray}\label{ap1.34}
&&\mathbb{P}_u\left(\|\delta\varphi\|_2^2>\varepsilon a\right)\le {\rm e}^{-\varepsilon ca}
\mathbb{E}_u\left\lbrack\exp\left( c\|\delta\varphi\|_2^2\right)\right\rbrack \nonumber \\
&&\le\left\lbrack C_3(\alpha)C_4(c)+C_6(\alpha)\, u^{1/2+g_2/2p-1/p} {\rm e}^{-\gamma(\varepsilon ,p)u}\right\rbrack\, 
{\rm e}^{-\varepsilon ca}.
\end{eqnarray}
Finally, using the estimates\ (\ref{ap1.17}) with $g_1=1$ and\ (\ref{ap1.34}) on the right-hand side of\ (\ref{eq3.2}) and taking the limit $u\rightarrow +\infty$, one obtains
\begin{equation}\label{ap1.35}
\lim_{u\rightarrow +\infty}\mathbb{P}_u\left(\|\delta\varphi\|_2^2>\varepsilon\|\overline{\varphi}\|_2^2\right)\le
 C_3(\alpha)C_4(c)\, {\rm e}^{-\varepsilon ca},
\end{equation}
and $a>0$ being arbitrary completes the proof of Proposition\ \ref{prop1} for a real field. [The right-hand side of\ (\ref{ap1.35}) can be made arbitrarily small.] $\square$
%
%
%
%

%
%
\end{document}